\documentclass[%
 aip,
 amsmath,amssymb,
 reprint,%
]{revtex4-1}

\usepackage[normalem]{ulem}
\usepackage{enumitem}%
\usepackage{amsmath}
\usepackage{multirow}
\usepackage{xcolor}
\usepackage{natbib}

\usepackage{graphicx}% Include figure files
\usepackage{dcolumn}% Align table columns on decimal point
\usepackage{bm}% bold math
\usepackage{float}
\usepackage[utf8]{inputenc}
\usepackage[T1]{fontenc}
\usepackage{mathptmx}
\usepackage{etoolbox}

\makeatletter
\def\@email#1#2{%
 \endgroup
 \patchcmd{\titleblock@produce}
  {\frontmatter@RRAPformat}
  {\frontmatter@RRAPformat{\produce@RRAP{*#1\href{mailto:#2}{#2}}}\frontmatter@RRAPformat}
  {}{}
}%
\makeatother
\begin{document}

\preprint{AIP/123-QED}

\title{An exploration of the phonon frequency spectrum and Born-von Karman periodic boundary conditions in 1D and 2D Lattice systems using a computational approach.}
% Force line breaks with \\
\author{J. Shannigrahi}
 
\author{P. Ashdhir}%
 \email{pragatiashdhir@hinducollege.ac.in}
\affiliation{ 
Department of Physics, Hindu College, University of Delhi%\\This line break forced with \textbackslash\textbackslash
}%

\date{\today}% It is always \today, today,

\begin{abstract}
Periodic boundary conditions (PBCS) are a pivotal concept in the treatment of ideal lattices of infinite extent as a finite lattice. Most undergraduate texts that delve into the analytical treatment of lattice dynamics do not explicitly incorporate the usage of PBCs. Moreover, most textbooks and existing literature predominantly solve the system in the frequency domain. The aim of the present work is to bridge this gap by demonstrating the application of Born von Karman PBCs in constructing a unit cell that effectively captures the dynamics of the entire lattice. The lattice dynamical equations are solved in the displacement-time domain using numerical methods. The Fast Fourier Transform technique is then used to obtain the phonon frequency spectrum corresponding to the computed instantaneous displacements. The approach explores the concept by first constructing the equations of atomic motion for linear lattices with and without a basis in the nearest neighbor approximation. Subsequently, such calculations are extended to obtain the phonon spectrum for two dimensional lattices such as the square lattice and the honeycomb lattice using the next-nearest neighbor approximations. A short range force constant model employing the central and angular forces is used for the monatomic square lattice. The dynamics of monatomic honeycomb lattice is investigated using the central forces to model the interatomic interactions. The computed results are validated against the analytical results of the given problem. Our work serves to showcase a novel method for understanding the implementation of PBCs in constructing a unit cell for a given lattice system and capturing its phonon dispersion spectrum. The approach is expected to be physically more intuitive for a student to understand the periodicity of a given lattice and its related dynamics. The target group of the present work comprise undergraduate students and educators who seek a thorough and didactic comprehension of lattice dynamics.

\end{abstract}

\maketitle

%%%%%%%%%%%%%%%%%%%%%%%%%%%%%%%%%%%%%%%%%%%%%%%%%%%%%%%%%%%%
\begin{figure*}
\centering
    \includegraphics{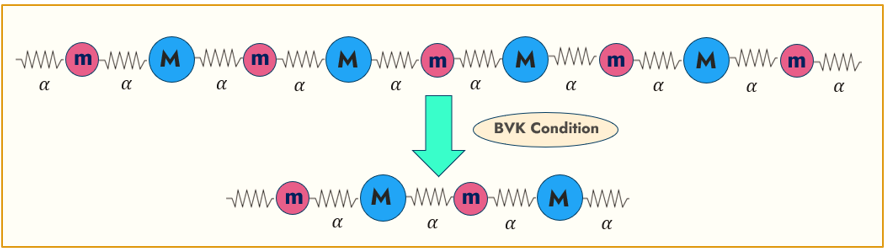}
    \caption{Implementation of Born von Karman periodic boundary conditions: An infinitely long linear diatomic lattice is condensed into an equivalent unit cell comprising of four atoms with two of each kind.}
    \label{fig:Figure BVK 1}
\end{figure*}

\section{\label{sec1:}Introduction:\protect }

The study of atomic vibrations in solids is of paramount importance in condensed matter physics. It provides useful insight for understanding the various physical properties of solids such as thermal conductivity \cite{Yang}, elasticity, thermal expansion coefficients and specific heat capacity \cite{M.Ali.Omar}$^{,}$\cite{Nasrollahi}. The arbitrary vibrational motion of the lattice can be considered a superposition of normal modes, such that each mode corresponds to the atoms oscillating uniformly at a specific frequency. The quanta of these lattice vibrations are known as phonons. The phonons are bosonic quasi-particles representing the collective excitations in crystalline solids. In analogy with the photons as quantized light waves, the phonons are quantized elastic waves propagating down a lattice \cite{Steven}.

In the existing vast wealth of literature and the reputed undergraduate level texts on solid state physics\cite{M.Ali.Omar}$^{,}$\cite{Steven}$^{,}$\cite{Patterson}$^{,}$\cite{Dekker}$^{,}$\cite{Ziman}$^{,}$\cite{HC_Gupta}$^{,}$\cite{Kittel}$^{,}$\cite{WA.Harris}$^{,}$\cite{Ashcroft}$^{,}$\cite{Kittel_Quant}$^{,}$\cite{RA_Levy}, the problem of lattice dynamics is traditionally solved in the reciprocal space rather than in the direct space. Typically, the analytical treatment comprises of deriving the dispersion relations for the phonon spectrum by making use of dynamical matrices\cite{Ashcroft}. These matrices contain all information about the dynamical behaviour of the crystal based on the by specific models for interatomic interactions.  The method intrinsically assumes the propagation of plane progressive waves\cite{Bloch} across a bounded lattice and consequently the existence of normal modes of atomic vibrations. This leads to the formulation of a secular determinant or characteristic equation yielding the normal mode frequencies of the given crystal system. Further, it is assumed that the end-effects or surface-effects have no bearing on the bulk properties of a crystalline solid\cite{Wallis}. This calls for the implementation of boundary conditions that allow us to treat an infinite lattice as a finite lattice. A convenient choice of boundary conditions for the running wave representation of lattice vibrational modes are the periodic boundary conditions proposed by Born and von Karman (BvK)\cite{M.Born}. An explicit implementation of Bvk PBCs in the lattice dynamical calculations is found missing in textbooks due to the cumbersome underlying mathematics.  

In this work, we propose a novel approach by numerically solving the lattice dynamical problem in direct space, which invites new opportunities for pedagogical discourse by explicitly emphasizing the implementation of periodic boundary conditions. It is expected to accord educators with a better tool to demonstrate the effect such conditions have on lattice structures.
Our method allows the solution of lattice dynamical equations in the physically more intuitive direct lattice-time domain vis-a-vis the reciprocal lattice-frequency domain. The dynamical equations for 1D and 2D lattice systems subjected to PBCs are solved using the fourth order Runge-Kutta method or the classic Runge-Kutta method as laid out in Kutta[1901]\cite{Kutta}. The phonon frequencies of the given lattice systems are obtained using the computational technique of Fast Fourier Transform (FFT). The computational approach as detailed in the subsequent sections attempts to  present an effective pedagogical technique for enhancing classroom discussions on lattice dynamics and the phonon spectrum. Python language is used as the computational platform for this work.  

In the following two subsections (\ref{sec1:lvl1} and \ref{sec1:lvl2}), we give a brief description of the requisites for understanding the methodology used in our approach to lattice dynamics.
\subsection{\label{sec1:lvl1}Model Formulation}
\begin{figure}[H]
    \centering
    \includegraphics[height=4cm,width=4cm]{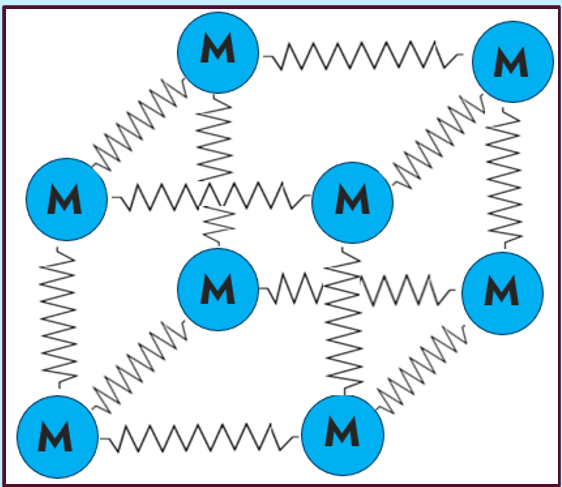}
    \caption{Illustration of harmonic approximations for a cubic lattice: Atoms are connected to each other with elastic springs that obey Hooke's law. }
    \label{fig:Harmonic}
\end{figure}

Lattice dynamics is the study of atomic vibrations in a crystal. A typical theory of lattice dynamics is based on the assumption that the equilibrium position of each constituent atom is a lattice site, about which it executes small oscillations. By small oscillations, it is implied that the displacement of atoms from their equilibrium is small compared with the interatomic spacing, so that the interatomic forces obey Hooke's law. This is equivalently the harmonic approximation under which only second order terms in atomic displacements are retained in the power series expansion of potential corresponding to the interatomic interactions. The atoms are hence modeled as being linked to each other through elastic springs \cite{Maradudin} as shown in Fig.\ref{fig:Harmonic}.

A crystal lattice is a periodic arrangement of atoms or groups of atoms in space. It may be generated by the spatial repetition of a unit cell containing some definite number of atoms. The purpose of the periodic boundary conditions is to allow us to construct a unit cell which simulates the behaviour of the entire lattice. An ideal lattice is considered to be effectively infinite in order to ignore the surface effects\cite{Wallis}.  The application of Born von Karman boundary conditions allow us to elegantly condense ideal lattices into a finite unit cell which exhibits phonon dispersion and other related properties, as the actual lattice of infinite extent\cite{M.Born}. 
The implementation of the Born von Karman conditions for an infinitely long linear diatomic lattice is illustrated in Fig.~\ref{fig:Figure BVK 1}. The lattice is condensed into a unit cell of four atoms, two of each kind, connected to each other by identical springs. The two dissimilar atoms at the two ends of the unit cell chain are assumed to be joined together so as to complete the loop. This way each atom in the unit cell has the atoms of the other kind as its immediate neighbors.

%%%%%%%%%%%%%%%%%%%%%%%%%%%%%%%%%%%%%%%%%%%%%%%%%%%
In the present work we have employed the Short Range Force Constant Model that takes into account only the short range interactions between the constituent atoms of a crystal. We have used both the central and angular type of short range forces for coupling the atoms, depending on the necessity borne by the complexity of the lattice structure. A detailed description of central and angular forces is given in section (\ref{sec5:}) The interatomic forces for only the nearest neighbors and next-nearest neighbors are taken into account as the higher order neighboring atoms are can be considered essentially screened in many practical applications\cite{Escande}. Similarly, even though the classical harmonic approximation is no longer valid in extremes of temperature\cite{Ashcroft}, it is a reasonable approximation for the purposes of this work. The coupled differential equations of motion so set-up for each atom in the condensed unit cell of a given lattice system are solved using the fourth order Runge-Kutta (RK-4) algorithm\cite{Kutta}.
%%%%%%%%%%%%%%%%%%%%%%%%%%%%%%%%%%%%%%%%%%%%%%%%%%%%%%%%%%%%
\subsection{\label{sec1:lvl2}Fast Fourier Transform Technique}
The Fast Fourier Transform (FFT) is a  popular computational technique that converts a time domain signal into individual spectral components and thereby provides frequency information about the signal. FFT is essentially an optimized algorithm for the implementation of the "Discrete Fourier Transformation" (DFT).

Our model is a novel illustration of the Fast Fourier Transform (FFT) algorithm in the physical sciences, diverging from its more common usage in signal processing. While a detailed overview of the method is beyond the scope of this work, interested readers can refer to the work by \textit{Ashdhir et al} for a more comprehensive explanation\cite{Ashdhir_etal}. We do, however, provide a concise description of the method as follows.

The FFT technique involves sampling a signal over a specific period of time. If the number of time domain samples is say, $N$, then implementation of the FFT algorithm on the sampled signal yields a $N$-point frequency spectrum. An accurate FFT implementation requires a wise choice of two critical parameters: the Sampling Rate (number of samples taken per unit time) and the Sampling Time Period (the total time duration for which the samples are recorded).  In the application of FFT to lattice vibrations, the computed instantaneous displacements of atoms serve as the displacement-time domain signals which are processed by the algorithm to yield the phonon frequency spectrum. In our case, a time-step of $2^{-16}$ s or $2^{-15}$ s is used to compute the time domain solutions of atomic displacements and velocities over a total time duration in the range $90-150$ s for the lattice systems considered. The particular choice of time-step and time-duration is arrived at by performing multiple runs of the code using different combinations of values for the two parameters to get accurate FFT results. The reciprocal of time step equal to $2^{16}$ s\textsuperscript{-1} or $2^{15}$ s\textsuperscript{-1} is the sampling rate and the time duration of $90-150$\ s is the sampling period for the given FFT computation. The FFT spectrum captures the normal mode frequencies at the high symmetry points within the first Brillouin zone (FBZ). The FBZ is a primitive unit cell in the reciprocal lattice space (also referred to as $k$-space) corresponding to the given real space lattice system. The periodicity of the lattice mandates that to obtain a unique relationship between the state of vibration of the lattice and the wave vector $\Vec{k}$, we need to confine the latter to within the FBZ. This justifies our approach on the analysis of the phonon spectrum within the first Brillouin zone.  

%%%%%%%%%%%%%%%%%%%%%%%%%%%%%%%%%%%%%%%%%%%%%%%%%%%%%%%%%%%%%%%%%%%%%%
\section{\label{sec2:}Monatomic Linear Chain\protect}
\begin{figure}[H] %1D Monatomic Chain Image
\centering
        \includegraphics[height=1.8cm,width=7.3cm]{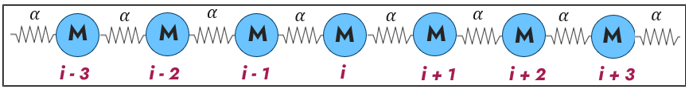}
        \caption{A segment of an infinitely long linear monatomic chain: Atoms of mass $M$ are connected to each other by elastic springs each of force constant $\alpha$.}
        \label{fig:1D Monatomic Lattice}
\end{figure}

\begin{figure}[H]
        \includegraphics[height=5cm,width=8.3cm]{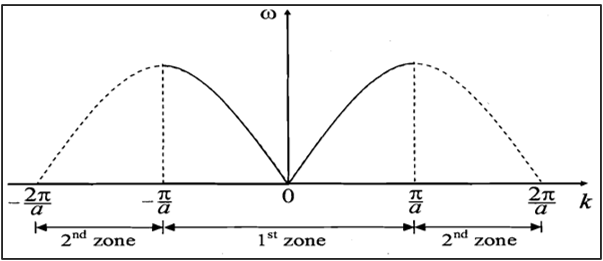}
        \caption{Dispersion relation for Monatomic Linear chain: Acoustical Branch extends from ${\omega=0}$ at the zone centre (${k=0}$) to ${\omega=2\ \sqrt{(\alpha/M)}}$ at the zone boundary (${k=\frac{\pi}{a}}$).}
        \label{fig:Dispersion Monatomic linear}
    \end{figure}

The monatomic linear chain is the simplest model for understanding the introductory lattice dynamics of a harmonic crystal. It consists of a one dimensional array of identical atoms connected by elastic bonds represented by springs each of force constant, say, $\alpha$. The equilibrium separation between two adjacent atoms is the lattice constant $a$. Figure ~\ref{fig:1D Monatomic Lattice} shows a segment of the infinitely long linear chain. When a longitudinal wave propagates through the monatomic linear chain, the constituent atoms get displaced from their respective equilibrium positions. Referring to the diagram, the equation of motion of the $i^{th}$ atom using the nearest neighbor approximation can be written as
\begin{equation}
\centering
    M \, \frac{d^2x_{i}}{{dt^2}} = -\alpha\ (x_{i}-x_{i+1})-\alpha\ (x_{i}-x_{i-1}),
    \label{eqn:0A}
\end{equation}
where $x_i$ is the instantaneous displacement of the $i^{th}$ atom from its equilibrium position and $M$ is the atomic mass. The first term on the RHS of the equation is the restoring force on the $i^{th}$ atom due to the $(i+1)^{th}$ atom and the second term is the restoring force on the $i^{th}$ atom due to the $(i-1)^{th}$ atom. In the traditional treatment, a plane wave solution of the form
\begin{equation}
\centering
    x_i =\zeta_i\ e^{(\Vec{k}.\Vec{r}-\omega t)} 
    \label{eqn:0B}
\end{equation}
is assumed for Eq.(\ref{eqn:0A}). Substituting Eq.(\ref{eqn:0B}) in Eq.(\ref{eqn:0A}), we get the dispersion relation for the linear monatomic lattice as
\begin{equation}
\centering
    \omega= 2\ \sqrt{\frac{\alpha}{M}}\ sin\big (\frac{ka}{2}\big ) 
    \label{eqn:0C}
\end{equation}
\hbadness=999999
As has been well documented in literature,\cite{M.Ali.Omar}$^{,}$\cite{Dekker}$^{,}$\cite{HC_Gupta}$^{,}$\cite{Kittel}$^{,}$\cite{Ashcroft} the monatomic linear lattice acts as a \textit{low-pass mechanical filter} such that the maximum phonon frequency that is allowed to pass through the lattice is equal to ${2\ \sqrt{(\alpha/M)}}$ at the Brillouin zone boundary ${k=\frac{\pi}{a}}$ where ${\alpha}$ is the central force constant and ${a}$ is the lattice constant. Higher frequencies will be strongly attenuated in the lattice. 
Figure~\ref{fig:Dispersion Monatomic linear} depicts the dispersion relation for a monatomic linear chain. It can be seen that the dispersion curve for a linear monatomic lattice has only the acoustical branch extending from ${\omega=0}$ at the zone centre (${k=0}$) to ${\omega=2\ \sqrt{(\alpha/M)}}$ at the zone boundary (${k=\frac{\pi}{a}}$). 

\begin{figure}[H] %1D Computational Model Image
    \centering
        \includegraphics[height=2.4cm,width=5.16cm]{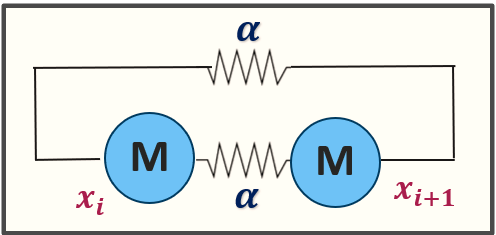}
        \caption{Computational Model for monatomic chain: Implementation of BvK periodic boundary conditions condenses the given infinite linear lattice into a lattice comprising of two identical atoms joined by two identical elastic springs in a cyclic manner.}
        \label{fig:Computational Model 1D Monatomic}
\end{figure}

Figure~\ref{fig:Computational Model 1D Monatomic} shows our computational model. By applying the BvK periodic boundary conditions consistent with nearest neighbor interactions, the given infinite linear lattice is reduced to a lattice comprising of two identical atoms joined by two identical elastic springs in a cyclical manner. If $x_i(t)$ and $x_{i+1}(t)$ are the instantaneous displacements of the two atoms, their equations of motion  can be written as:
\begin{equation}
\centering
    M \, \frac{d^2x_{i}}{{dt^2}}= - {\alpha(2\,\,x_{i}-x_{i+1}-x_{i+1})}
    \label{eqn:0D}
\end{equation}
\begin{equation}
\centering
    M \, \frac{d^2x_{i+1}}{{dt^2}} =  - {\alpha(2\,\,x_{i+1}-x_{i}-x_{i})}
    \label{eqn:0E}
\end{equation}
The given coupled second order differential equations are numerically solved using the fourth order Runge Kutta method\cite{Kutta} under different sets of initial conditions. 

%%%%%%%%%%%%%%%%%%%%%%%%%%%%%%%%%%%%%%%%%%%%%%%%%%%%%%%%%%%%
%%%%%%%%%%%%%%%%%%%%%%%%%%%%%%%%%%%%%%%%%%%%%%%%%%%%%%%%%%%%%%%%%%%%%%%%%%%%%%
 \begin{figure}[h]%1D Monatomic FFT 1 Image 1
        \includegraphics[height=7cm,width=7.8cm]{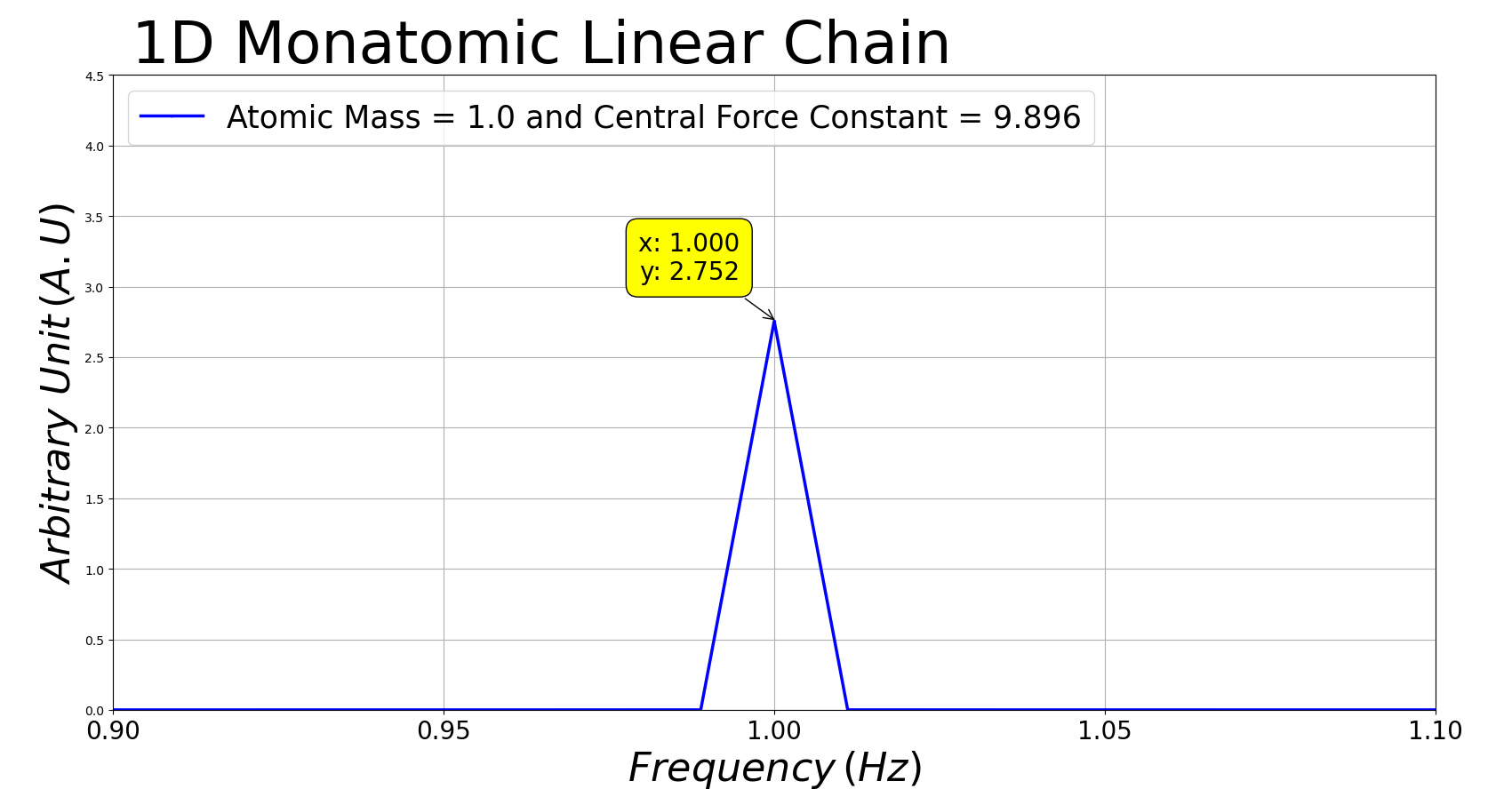}
        \caption{ FFT plot for a monatomic linear chain with $m=1$ and $\alpha=\pi^2$: Observed frequency peak at $1.000$ with initial displacements of the atoms being $x_{1}=1$, $x_{2}=7$ and $\Dot{x_1}=\Dot{x_2}=0$.}
        \label{fig:Monatomic linear FFT 1 Image 1}
    \end{figure}
    
    \begin{figure}[h] %1D Monatomic FFT 1 Image 2
        \includegraphics[height=7cm,width=7.8cm]{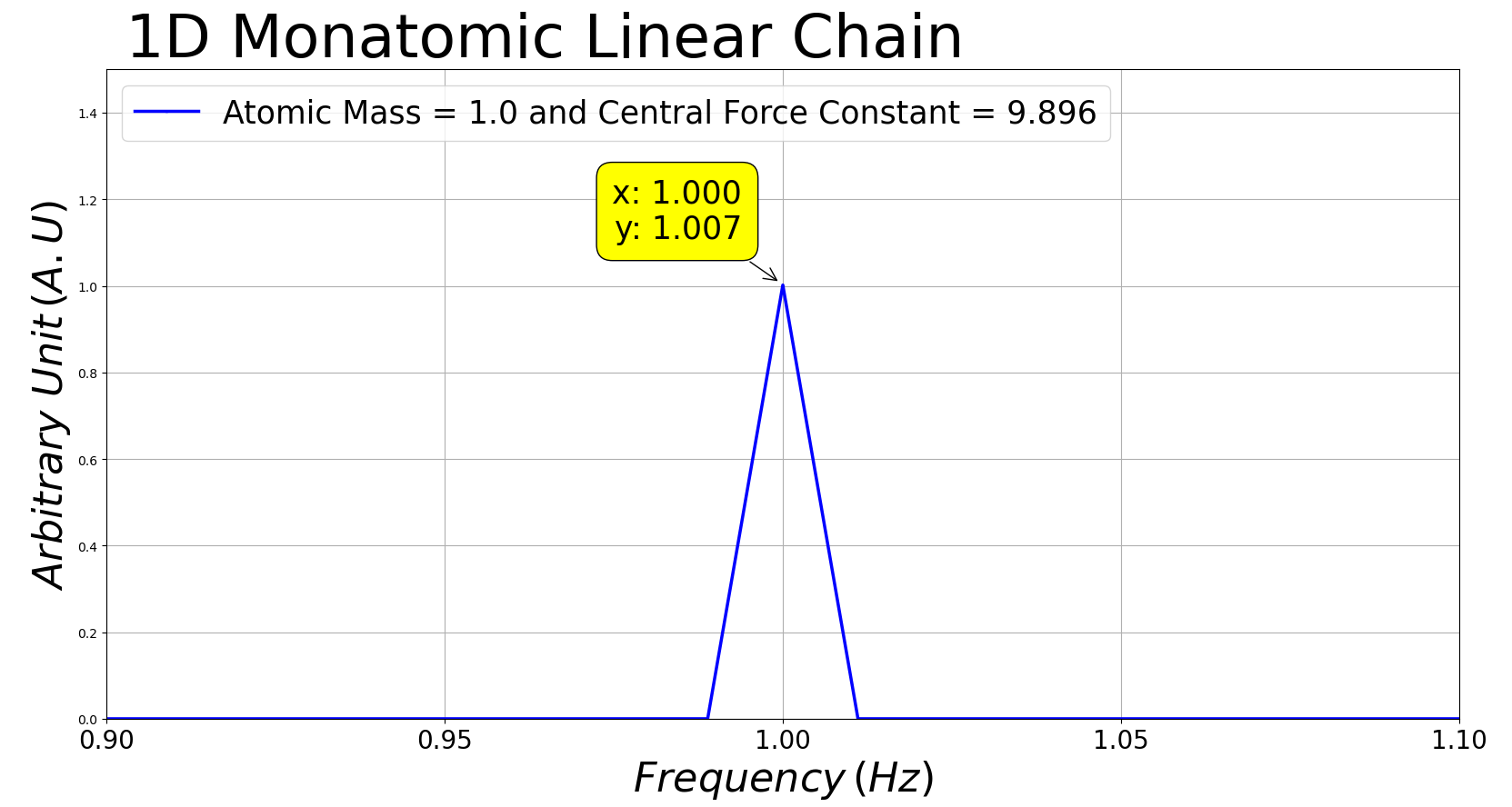}
        \caption{ FFT plot for a monatomic linear chain with $m=1$ and $\alpha=\pi^2$: Observed frequency peak at $1.000$ with initial displacements of the atoms being $x_{1}=2.5$, $x_{2}=4.5$ and $\Dot{x_1}=\Dot{x_2}=0$.}
        \label{fig:Monatomic linear FFT 1 Image 2}
    \end{figure}

        \begin{figure}[h]%1D Monatomic FFT 2 Image 1
        \includegraphics[height=7cm,width=7.8cm]{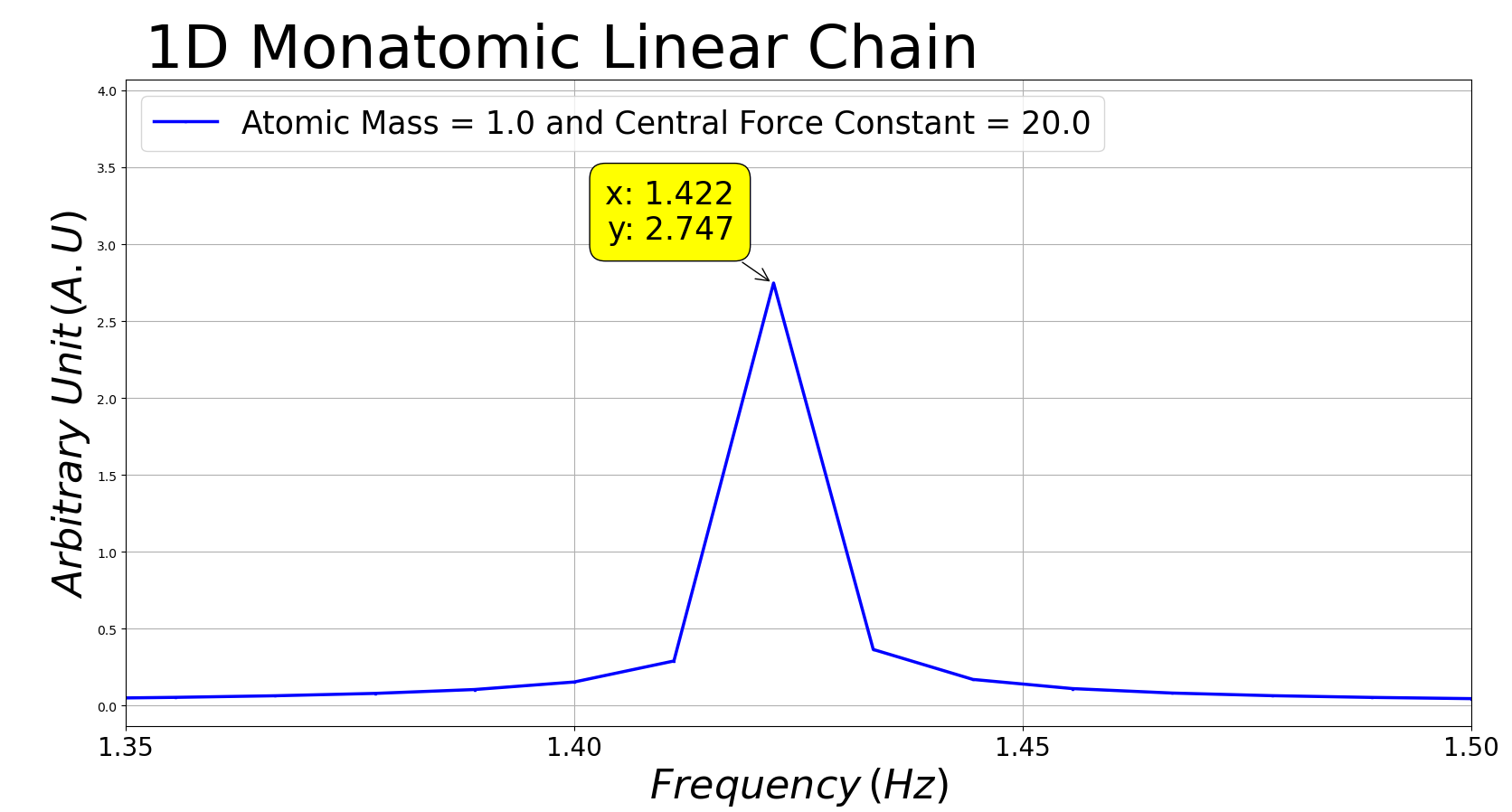}
        \caption{ FFT plot for a monatomic linear chain with $m=1$ and $\alpha=20$: Observed frequency peak at $1.422$ with initial displacements of the atoms being $x_{1}=1$, $x_{2}=8$ and $\Dot{x_1}=\Dot{x_2}=0$.}
        \label{fig:Monatomic linear FFT 2 Image 1}
    \end{figure}
    
    \begin{figure}[h] %1D Monatomic FFT 2 Image 2
        \includegraphics[height=7cm,width=7.8cm]{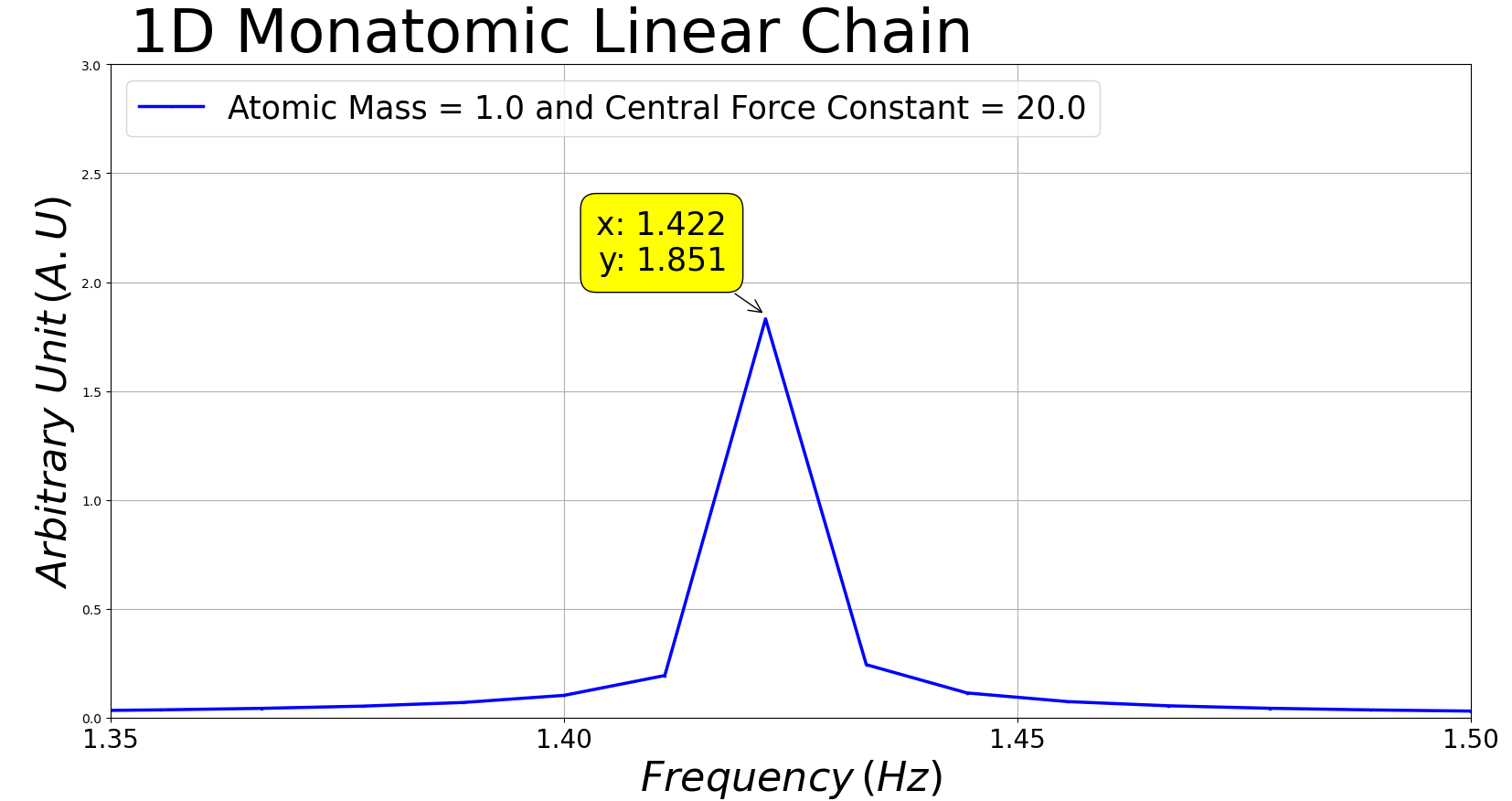}
        \caption{FFT plot for a monatomic linear chain with $m=1$ and $\alpha=20$: Observed frequency peak at $1.422$ with initial displacements of the atoms being $x_{1}=-1$, $x_{2}=3$ and $\Dot{x_1}=\Dot{x_2}=0$.}
        \label{fig:Monatomic linear FFT 2 Image 2}
    \end{figure}
%%%%%%%%%%%%%%%%%%%%%%%%%%%%%%%%%%%%%%%%%%%%%%%%%%%%%%%%%%%%%%%%%%%%%%%%%%%%%%

The FFT algorithm is hence applied to the time-domain computed solutions of equations Eq.(\ref{eqn:0D}) and Eq.(\ref{eqn:0E}). The applicability of the algorithm is verified in two independent cases (${\alpha=\ \pi^2,\ 20}$ \& $m=1$) with different sets of initial conditions. It is found that the analytical and FFT computed values of the phonon frequency at the FBZ boundary have a very good agreement as shown in Table \ref{table:1D Monatomic FFT table} : 
\begin{table}[h!]
    \centering
    \renewcommand{\arraystretch}{1.2} % Adjust vertical spacing
      \begin{tabular}{|>{\centering\arraybackslash}m{1.0 cm}|>{\centering\arraybackslash}m{1.0 cm}|>{\centering\arraybackslash}m{1.5 cm}|>{\centering\arraybackslash}m{1.5 cm}|>{\centering\arraybackslash}m{2 cm}|}
          \hline
           \multirow{2}{3em}{Case No.} & \multirow{2}{3em}{Atomic Mass} & \multirow{2}{4em}{Spring Constant} & \multirow{2}{5em}{Theoretical Value} & \multirow{2}{7em}{FFT Computed Value}\\ [7pt]
          &&&&\\
          \hline
            1. & $1.0$ &  $\pi^{2}$ & $1.000$ &  
            \textcolor{red}{$1.000$}\\[7pt]
          \hline
            2. & $1.0$ &  $20.0$ & $1.423$ &  
            \textcolor{red}{$1.422$}\\[7pt]
          \hline
      \end{tabular}
      \caption{Comparison of theoretical and FFT computed values of phonon frequency at the first Brillouin zone boundary for a monatomic linear chain.}
    \label{table:1D Monatomic FFT table}
  \end{table}

The FFT plots of the given two cases are depicted in Fig.~\ref{fig:Monatomic linear FFT 1 Image 1} through Fig.~\ref{fig:Monatomic linear FFT 2 Image 2}. It is evident from the plots that the application of FFT effectively captures the phononic behaviour of the system at the FBZ boundary, irrespective of the initial excitation energies of the atoms. As is expected, we observe that the computed frequency is independent of the initial energy of the atoms. The initial excitation energy of the atoms only affect the height of the FFT peaks, which are indicative of the amplitudes of lattice vibrations corresponding to that specific frequency. 

\section{\label{sec3:}Diatomic Linear Chain with a basis \protect }
\begin{figure}[h] %1D Diatomic Chain
        \centering
        \includegraphics[height=2.5cm,width=8.0cm]{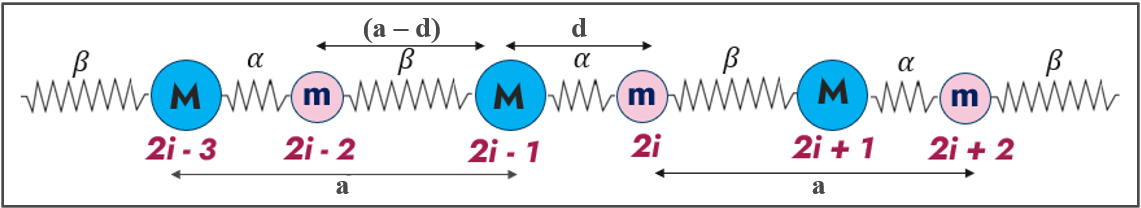}
        \caption{A Diatomic Linear Chain with a Basis: It comprises of two types of atoms with different masses such that the similar atoms are separated by distance $a$ and the dissimilar atoms are separated by distances $d$ and $(a-d)$.}
        \label{fig:1D Diatomic chain}
    \end{figure}

The diatomic linear chain represents a bit more complex case of modeling of a lattice structure. It is a lattice with dissimilar atoms of masses, say, \textit{m} and \textit{M} assuming \textit{(M>m)}. The similar atoms are separated by distance $a$ and the dissimilar atoms are separated by distances $d$ and $(a-d)$ as shown in Fig.~\ref{fig:1D Diatomic chain}. Hence, two springs constants $\bm{\alpha}$ and $\bm{\beta}$ are considered between the masses \textit{m} and \textit{M}. It can be viewed as a non-Bravais lattice comprising of two interpenetrating Bravais lattices, each corresponding to a given atom type. When a longitudinal wave passes through the given lattice, the atoms get displaced from their respective equilibrium positions and the equations of motion in the nearest neighbor approximation can be formulated as
\begin{equation}
\centering
    m \, \frac{d^2x_{2\,i}}{{dt^2}} = - {\beta(x_{2\,i}-x_{2\,i+1}) - \alpha(x_{2\,i}-x_{2\,i-1})}
    \label{eqn:1A}
\end{equation}
\begin{equation}
\centering
    M \, \frac{d^2x_{2\,i+1}}{{dt^2}} = - {\beta(x_{2\,i+1}-x_{2\,i}) - \alpha(x_{2\,i+1}-x_{2\,i+2})}
    \label{eqn:1B}
\end{equation}
Once again, the traditional treatment assumes the existence of plane wave solutions of the form given in Eq.(\ref{eqn:0B}) to satisfy the above two equations of motion yielding the dispersion relation 

\begin{figure}[H] %1D Diatomic Chain Dispersion relation
    \centering
        \includegraphics[height=5.5cm,width=6.5cm]{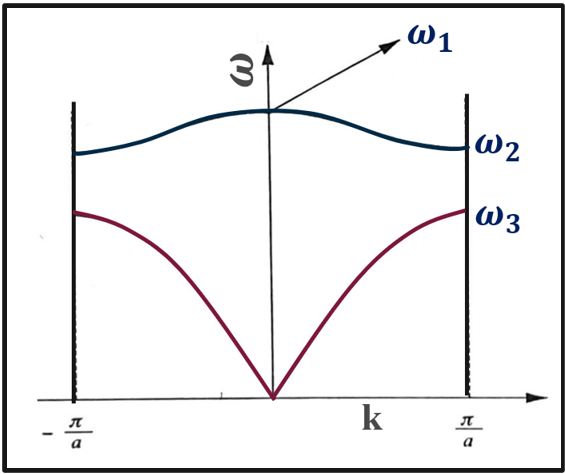}
        \caption{Phonon Dispersion Curve for Diatomic Chain with Basis: The curve has two branches, one acoustical and one optical branch. The upper optical branch frequencies decrease from the zone center value $\omega_{1} = {\sqrt{\frac{{(\alpha + \beta)(m + M)}}{{m\,M}}}} $ to the zone boundary value $\omega_{2} = \sqrt{{\frac{{\alpha + \beta}}{{m}}}}$. The lower acoustic branch rises from zero at the zone center to zone boundary value $\omega_{3} = \sqrt{{\frac{{\alpha + \beta}}{{M}}}}$.}
        \label{fig:1D Diatomic Dispersion relation}
    \end{figure}

\begin{multline}
     \omega^{2}=\frac{(\alpha+\beta)(m+M)}{2\ m\ M} \\ 
     \pm\frac{\sqrt{(m+M){^2}(\alpha+\beta)^{2}-8 m M \alpha \beta (1-cos(k a)}}{2\ m\ M}   
    \label{eqn:1C}
\end{multline}
for the given diatomic lattice with a basis. Figure~\ref{fig:1D Diatomic Dispersion relation} depicts the above dispersion relation. It can be seen that for the diatomic chain with a basis, the dispersion curve has two frequency branches associated with it. The upper dispersion branch known as the \textit{optical branch} is non-zero at $k=0$, the zone center frequency being given by Eq.(\ref{eqn:1}).
\begin{equation}
    \centering
    \omega_{1} = {\sqrt{\frac{{(\alpha + \beta)(m + M)}}{{m\,M}}}}  
    \label{eqn:1}
\end{equation}
The optical branch frequencies decrease from the zone center value (Eq.(\ref{eqn:1})) to the value at the FBZ boundary given by Eq.(\ref{eqn:2}).
\begin{equation}
    \centering
    \omega_{2} = \sqrt{{\frac{{\alpha + \beta}}{{m}}}}
    \label{eqn:2}
\end{equation}
The lower branch of the dispersion curve is the \textit{acoustic branch}. It rises from zero at the zone center to a value given by Eq.(\ref{eqn:3}) at the FBZ boundary.
\begin{equation}
    \centering
    \omega_{3} = \sqrt{{\frac{{\alpha + \beta}}{{M}}}} \,\,\,\,\,\,\, \, (for \,\,\, m<M),
    \label{eqn:3}
\end{equation}
\noindent where $\omega_{1},\ \omega_{2},\ \omega_{3} $ represent the angular frequencies. The corresponding temporal frequencies $ f_{1}, \ f_{2}, \ f_{3}$ can be found using the relation $f = \frac{1}{2\pi}\,\omega$. This shows that a diatomic chain behaves as a \textit{band-pass mechanical filter} since the  frequencies corresponding to those between the acoustic and optical branches at the zone boundary (Eq.(\ref{eqn:2}) and Eq.(\ref{eqn:3}) ) are forbidden by the system\cite{M.Ali.Omar}.
 \begin{figure}[H] %2D Diatomic Computational model
        \centering
        \includegraphics[height=3.0cm,width=7.5cm]{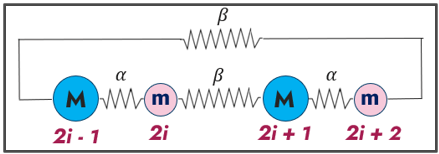}
        \caption{Computational model of a Diatomic Chain with Basis: Implementation of BvK periodic boundary conditions condenses the given infinite linear lattice into a lattice comprising of four atoms, two of each kind joined by two kinds of elastic springs in a cyclic manner.}
        \label{fig:Computational model of diatomic chain}
    \end{figure} 
     
The computational model of the diatomic chain  obtained after implementation of BvK periodic boundary conditions is shown in Fig.~\ref{fig:Computational model of diatomic chain}. It comprises of four atoms, two of each kind. The chain unit is completed by connecting the $(2i+2)^{th}$ atom of mass $m$ with the $(2i-1)^{th}$ atom of mass $M$. The condensation of an infinitely long diatomic lattice with basis into a finite unit of four atoms can also be understood in terms of interpenetration of two condensed monatomic lattices similar to the one depicted in Fig.~\ref{fig:Computational Model 1D Monatomic}, one each for masses $m$ and $M$ with force constants $\alpha$ and $\beta$ respectively. In the nearest neighbor approximation, the coupled equations of atomic motion can be written as 
\begin{equation}
\centering
    m \, \frac{d^2x_{2\,i-1}}{{dt^2}} = - {\alpha(x_{2\,i-1}-x_{2\,i+2}) - \beta(x_{2\,i-1}-x_{2\,i})}
    \label{eqn:1D}
\end{equation}

\begin{equation}
\centering
    M \, \frac{d^2x_{2\,i}}{{dt^2}} = -{\alpha(x_{2\,i}-x_{2\,i-1}) - \beta(x_{2\,i}-x_{2\,i+1})}
    \label{eqn:1E}
\end{equation}

\begin{equation}
\centering
    m \, \frac{d^2x_{2\,i+1}}{{dt^2}} = -{\alpha(x_{2\,i+1}-x_{2i}) - \beta(x_{2\,i+1}-x_{2\,i+2})}
    \label{eqn:1F}
\end{equation}

\begin{equation}
\centering
    M \, \frac{d^2x_{2\,i+2}}{{dt^2}} = -{\alpha(x_{2\,i+2}-x_{2\,i+1}) - \beta(x_{2\,i+2}-x_{2\,i-1})}
    \label{eqn:1G}
\end{equation}
The equations are solved using the Fourth Order Runge Kutta method to obtain the numerical solutions for instantaneous positions and velocities of the atoms. The choice of initial conditions for atomic displacements and velocities is randomized (using NumPy's \textit{random} module in Python).

%%%%%%%%%%%%%%%%%%%%%%%%%%%%%%%%%%%%%%%
 \begin{figure}[H]%1D Diatomic FFT Image 
        \centering
        \includegraphics[height=7.8cm,width=8.6cm]{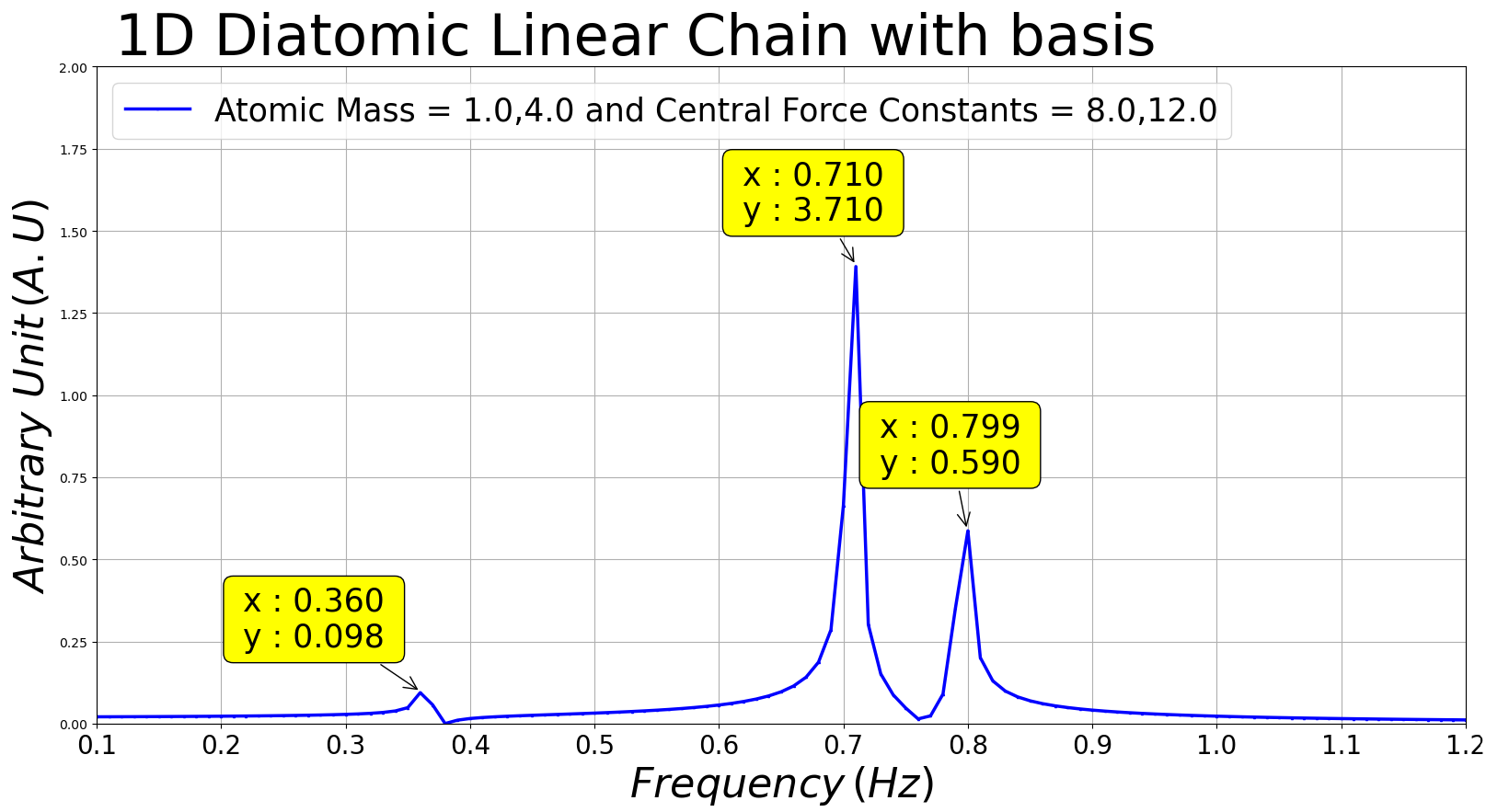}
        \caption{FFT Plot for an atom of a diatomic chain with a basis exhibits $3$ peaks: Two peaks are mapped with the two optical phonon modes and one with the acoustical phonon mode}
        \label{fig:Diatomic linear}
    \end{figure}
The FFT of the computed instantaneous displacements for each of the four atoms in the condensed lattice unit of the successfully captures the three phonon frequencies, two corresponding to the FBZ boundary and  one to the zone center. The atomic masses and force constants in arbitrary units are taken as: $m = 1.0  \,\,; \,\, M = 4.0 \,\,; \,\, \alpha = 8.0 ; \,\, \beta = 12.0$. The FFT plot for only one of the four atoms is shown in Fig.~\ref{fig:Diatomic linear}. The corresponding plots of the other three atoms are not shown because as expected, those atoms also exhibit three peaks at identical frequencies, two of which can be mapped with the two optical phonon modes ($\omega_1$ \& $\omega_2$) and one with the acoustical phonon frequency ($\omega_3$). The magnitudes of FFT peaks are insignificant in the context of present study. The height of the peaks are determined by the initial excitation conditions imposed on the lattice, which in our case are randomized to eliminate any prejudices in our computed phonon frequencies. Physically, the magnitude of FFT peaks are indicative of amplitude of atomic vibrations in a given normal mode of the lattice.
\begin{table}[h!]
    \centering
      \renewcommand{\arraystretch}{1.2} % Adjust vertical spacing
      \begin{tabular}{|>{\centering\arraybackslash}m{1 cm}|>{\centering\arraybackslash}m{1 cm}|>{\centering\arraybackslash}m{1 cm}|>{\centering\arraybackslash}m{1 cm}|>{\centering\arraybackslash}m{1 cm}|>{\centering\arraybackslash}m{1 cm}|>{\centering\arraybackslash}m{1 cm}|>{\centering\arraybackslash}m{1 cm}|}
          \hline
             \multicolumn{3}{|c|}{Theoretical Values} & \multicolumn{3}{|c|}{FFT Values}  \\[10pt]
          \hline
          $f_{1}$ & $f_{2}$ & $f_{3}$ & $f_{1}^{'}$ & $f_{2}^{'}$ & $f_{3}^{'}$ \\[10pt]
          \hline
            $0.795$&  $0.712$&  $0.353$
          &\textcolor{red}{$0.799$} &  \textcolor{red}{$0.710$} & \textcolor{red}{$0.360$} \\[12pt]
          \hline
      \end{tabular}
    \caption{Comparison of theoretical and FFT computed values of zone centre and zone boundary phonon frequencies for a diatomic chain with a basis.}
    \label{table:1D Diatomic with basis FFT table}
  \end{table}
Table \ref{table:1D Diatomic with basis FFT table} gives the theoretical and computed values of the zone centre and zone boundary phonon frequencies. There is a close agreement between the two sets of values within a maximum absolute error of around {$2\%$}. This indicates the accuracy and fidelity of the model's predictions.
%%%%%%%%%%%%%%%%%%%%%%%%%%%%%%%%%%%%%%%%%%%%%%%%%%%%%%%%%%%%%%%%%%%%%
\section{\label{sec4:lvl1}Diatomic Chain without a Basis}
\begin{figure}[H] %2D Diatomic chain without basis
        \centering
        \includegraphics[height=3.2cm,width=8.0cm]{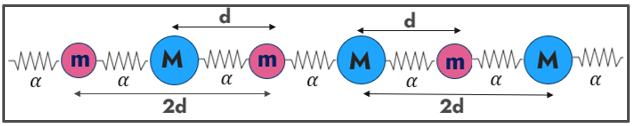}
        \caption{Diatomic Linear Chain: Infinitely long chain of two dissimilar atoms of masses $m$ and $M$ at equal separation $d$ joined by identical elastic springs with force constant $\alpha$.}
        \label{fig:1D Diatomic chain without basis}
    \end{figure}

The diatomic linear chain without a basis is a special case of the  lattice system described in section (\ref{sec3:}). It is a Bravais lattice with two dissimilar atoms of masses $m$ and $M$ placed alternately at identical separation distances. The atoms are coupled to each other by identical elastic springs with force constant $\alpha$ as depicted in Fig.~\ref{fig:1D Diatomic chain without basis}.

The analytical expressions for phonon frequencies of the given lattice can be easily arrived at by putting $\alpha=\beta$ in Eq.(\ref{eqn:1})- Eq.(\ref{eqn:3}) to yield the   

Optical branch frequency at zone centre as
\begin{equation}
    \centering
        \omega_{1} = {\sqrt{\frac{{(2\,\alpha)(m + M)}}{{m\,M}}}},
    \label{eqn:4}
\end{equation}

Optical branch frequency at zone boundary as
\begin{equation}
    \centering
        \omega_{2} = \sqrt{{\frac{{(2\,\alpha)}}{{m}}}}, 
    \label{eqn:5}
\end{equation}

Acoustic branch frequency at zone boundary:
\begin{equation}
    \centering
        \omega_{3} = \sqrt{{\frac{{(2\,\alpha)}}{{M}}}} \,\,\,\,\,\,\, \, (for \,\,\, m<M). 
    \label{eqn:6}
\end{equation}

The condensed lattice unit after applying the BvK PBCs comprises of four atoms as depicted in Fig.~\ref{fig:1D Diatomic chain without basis}. It is similar to the earlier case in Fig.~\ref{fig:Computational model of diatomic chain}, except that the force constant $\beta$ is replaced by $\alpha$. 
The corresponding equations of motion are also similar in form to Eq.(\ref{eqn:1D}) through Eq.(\ref{eqn:1G}) with $\alpha = \beta$. Similar to the procedure in previous cases, the FFT algorithm is applied to the computed time domain displacement solutions of the newly formulated equations of motion. The computational parameters are taken as: $m = 1.0  \,\,; \,\, M = 4.0 \,\, \& \,\, \alpha = 10.0 $. 

    \begin{figure}[H]%1D Diatomic without basis FFT Image 1
        \centering
        \includegraphics[height=6.8cm,width=8.0cm]{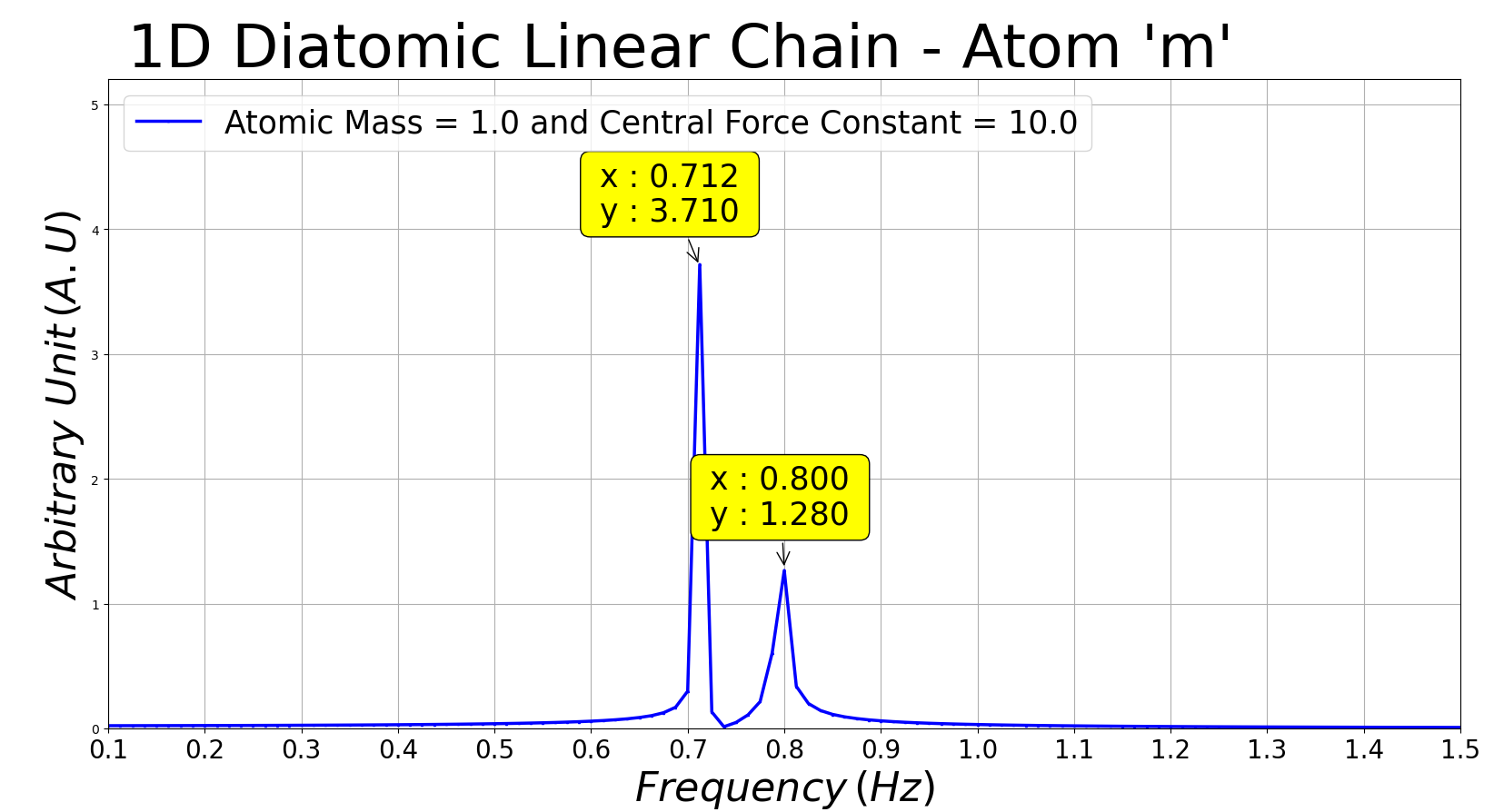}
        \caption{FFT plot for the lighter atoms with mass $m$ of a diatomic chain without basis: Two peaks are mapped with $\omega_{1} = {\sqrt{\frac{{(2\,\alpha)(m + M)}}{{m\,M}}}}$ and $\omega_{2} = \sqrt{{\frac{{(2\,\alpha)}}{{m}}}}$.}
        \label{fig:Diatomic chain without basis 1}
    \end{figure}

    \begin{figure}[H]%1D Diatomic without basis FFT Image 2
        \centering
        \includegraphics[height=6.8cm,width=8.0cm]{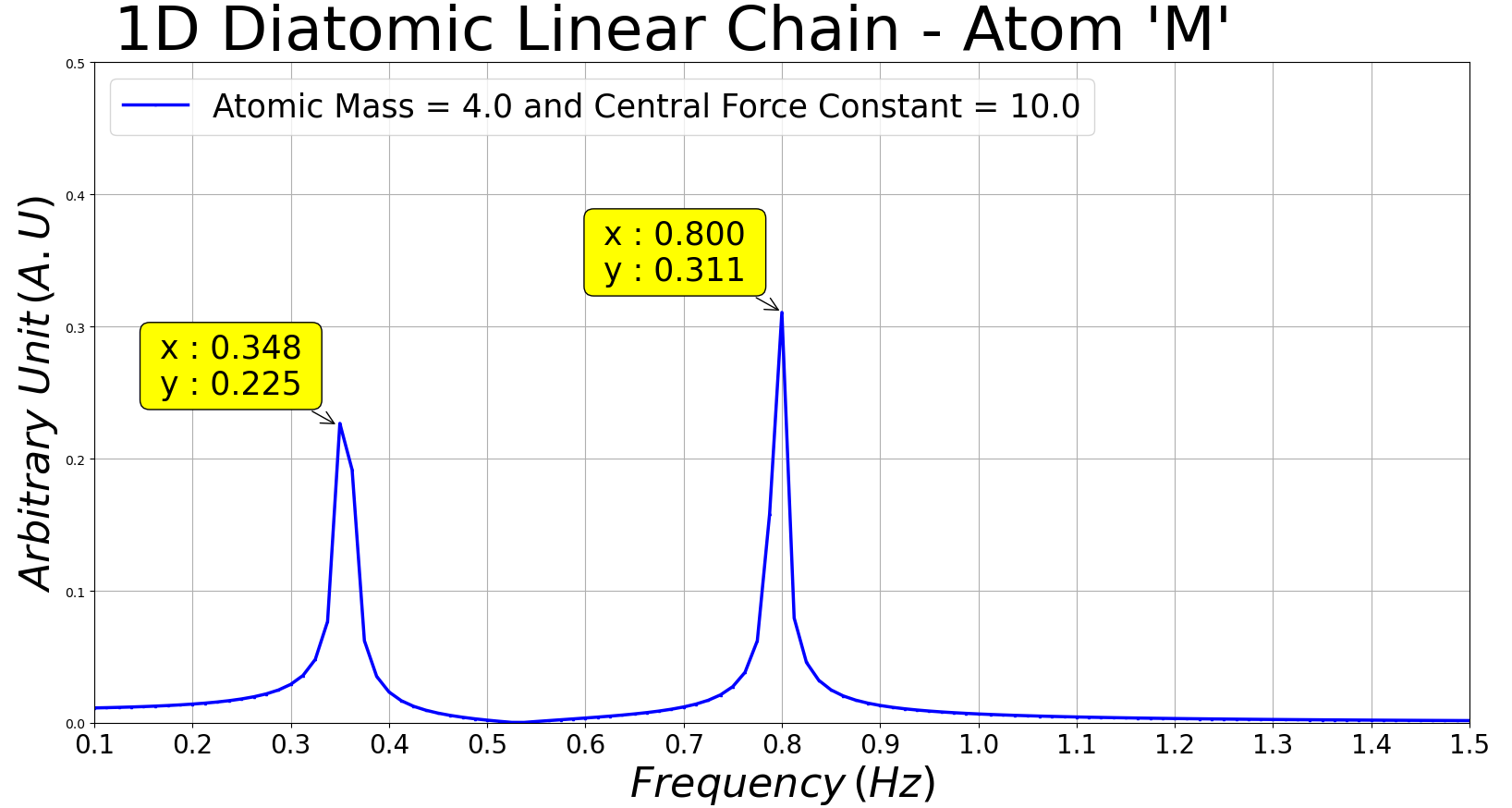}
        \caption{ FFT plot for the heavier atoms with mass $M$ of a diatomic chain without basis: Two peaks are mapped with  $\omega_{1} = {\sqrt{\frac{{(2\,\alpha)(m + M)}}{{m\,M}}}}$ and $\omega_{3} = \sqrt{{\frac{{(2\,\alpha)}}{{M}}}}$.}
        \label{fig:Diatomic chain without basis 2}
    \end{figure}

Interestingly, this time the normal mode frequencies as captured by FFT turn out to be different for the two kinds of atoms as depicted in Fig.~\ref{fig:Diatomic chain without basis 1} and Fig.~\ref{fig:Diatomic chain without basis 2}. Each of the atom types  exhibit equal frequencies at $0.800$ Hz corresponding to the optical branch at the zone center ( $k=0$) given by Eq.(\ref{eqn:4}). However they differ in phonon frequency peaks at the zone boundary, the lighter atoms with mass $m$ exhibit a peak at $0.712$ Hz given by Eq.(\ref{eqn:5}), while the the heavier atoms with mass $M$ exhibit a peak at at $0.348$ Hz given by Eq.(\ref{eqn:6}). It can thus be concluded that while both types of atom participate in the optical normal mode of vibration, the acoustical mode of vibration involves only the heavier atoms.
\begin{table}[h!]
    \centering
    \renewcommand{\arraystretch}{1.2} % Adjust vertical spacing
      \begin{tabular}{|>{\centering\arraybackslash}m{1 cm}|>{\centering\arraybackslash}m{1 cm}|>{\centering\arraybackslash}m{1 cm}|>{\centering\arraybackslash}m{1 cm}|>{\centering\arraybackslash}m{1 cm}|>{\centering\arraybackslash}m{1 cm}|>{\centering\arraybackslash}m{1 cm}|>{\centering\arraybackslash}m{1 cm}|}
          \hline
             \multicolumn{3}{|c|}{Theoretical Values} & \multicolumn{3}{|c|}{FFT Values}  \\[10pt]
          \hline
          $f_{1}$ & $f_{2}$ & $f_{3}$ & $f_{1}^{'}$ & $f_{2}^{'}$ & $f_{3}^{'}$ \\[10pt]
          \hline
            $0.795$&  $0.712$&  $0.353$
          &\textcolor{red}{$0.800$} &  \textcolor{red}{$0.712$} & \textcolor{red}{$0.348$} \\[12pt]
          \hline
      \end{tabular}
    \caption{Comparison of theoretical and FFT computed values of zone centre and zone boundary phonon frequencies for a diatomic chain without a basis.}
    \label{table:1D Diatomic FFT table}
  \end{table}

The computed and theoretical values of the FBZ phonon frequencies for the given diatomic lattice without basis have a fairly good agreement as shown in Table~\ref{table:1D Diatomic FFT table}. Our computational model explicitly highlights the subtly distinct behaviour of  diatomic lattices with and without basis with regard to the normal modes of vibration.  
%%%%%%%%%%%%%%%%%%%%%%%%%%%%%%%%%%%%%%%%%%%%%%%%%%%%%%%%%%%%%%%%%%%%%%%%%%%%%%%%%%%%%%%%%%%%%%%
\section{\label{sec5:}Monatomic Square Lattice}

    \begin{figure}[H]%2D Square Lattice
        \centering
        \includegraphics[height=6.5cm,width=6.5cm]{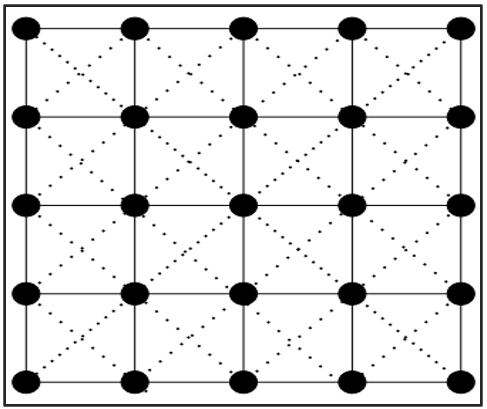}
        \caption{Monatomic square lattice structure: Solid lines connect the nearest neighbors and the dashed lines connect the second nearest neighbors.}
        \label{fig:Monatomic Square lattice}
    \end{figure}
    
The monatomic square lattice represents the simplest case of modeling a two-dimensional lattice structure. It is a Bravais lattice with identical atoms arranged in a planar pattern shown in Fig.~\ref{fig:Monatomic Square lattice}. Each of the atoms in this lattice structure has two degrees of freedom which somewhat increases the complexity of lattice dynamics.

To work out the dynamics of lattices with more than one dimension we may need to model the interatomic interactions by non-central or angular forces in addition to the central forces. The two-body central forces that we considered for 1D lattice systems were assumed to act in a direction collinear with the equilibrium line connecting the two interacting atoms and to arise out of the instantaneous relative displacement between them. The two-body non-central or angular forces as defined by de Launey\cite{de_launey} depend on the angle which the line joining the moving atoms makes with the equilibrium position of the line.
\begin{figure}[H]%Angular force illustration
        \centering
        \includegraphics[height=6.5cm,width=6.5cm]{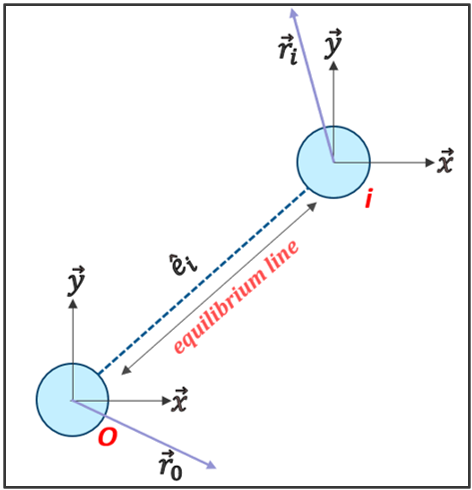}
        \caption{Illustration of Central and deLauney type Angular forces to model interatomic interactions.}
        \label{fig:angular force illustrated}
\end{figure}
Using the illustration in Fig.~\ref{fig:angular force illustrated}, the analytical expressions for central and deLauney type of angular forces can be written as
\begin{equation}
    \centering
    \vec{\mathbf{F}}_{\text{central}} = -\alpha\,[\, \hat{\mathbf{e}}_{i} \, \cdot \, (\vec{\mathbf{r}}_{o} - \vec{\mathbf{r}}_{i}) ] \, \cdot \hat{\mathbf{e}}_{i}
    \label{eqn:7}
\end{equation}
\begin{equation}
    \centering
    \vec{\mathbf{F}}_{\text{angular}} = -\beta\,[\, \hat{\mathbf{e}}_{i} \, \times \, (\vec{\mathbf{r}}_{o} - \vec{\mathbf{r}}_{i}) ] \, \times \hat{\mathbf{e}}_{i},
    \label{eqn:8}
\end{equation}
where \textbf{\emph{o}} represents the reference atom and \textbf{\emph{i}} represents a neighboring atom ; $\hat{\mathbf{e}}_{i}$ is a unit vector in the direction of the equilibrium line joining the two atoms ; $\vec{\mathbf{r}}_{i}$ and $\vec{\mathbf{r}}_{o}$ are the instantaneous displacements of the $i^{th}$ and $o^{th}$ atoms respectively; $\alpha$ and $\beta$ are the central and angular force constants respectively.\\
The net restoring force on atom \textbf{\emph{o}} due to its interaction with atom \textbf{\emph{i}} is given by the vector sum of the central and angular forces as
\begin{equation}
    \centering
    \vec{\mathbf{F}} = \vec{\mathbf{F}}_{\text{central}}+\vec{\mathbf{F}}_{\text{angular}}
    \label{eqn:8A}
\end{equation}
\begin{equation}
    \centering
    \vec{\mathbf{F}} = -\beta\,(\vec{\mathbf{r}}_{o} - \vec{\mathbf{r}}_{i})-(\alpha-\beta)\,[\, \hat{\mathbf{e}}_{i} \, \cdot \, (\vec{\mathbf{r}}_{o} - \vec{\mathbf{r}}_{i}) ] \, \cdot \hat{\mathbf{e}}_{i}
    \label{eqn:8B}
\end{equation}
The resultant force as given in Eq.(\ref{eqn:8B}) can be resolved into the Cartesian components \textbf{\emph{x}}, \textbf{\emph{y}} as given below:
\begin{center}
    \begin{multline}
    {F}_{x} = -\beta\,( {{x}}_{o} -  {{x}}_{i})-(\alpha-\beta)\,  {{e}_{ix}} \,  \,[\,   {{e}}_{ix} \,  \, ( {{x}}_{o} -  {{x}}_{i})  \\+    {{e}_{iy}} \,  \, ( {{y}}_{o} -  {{y}}_{i}) ]
    \label{eqn:8C}
\end{multline}
\end{center}
\begin{center}
    \begin{multline}
    {{F}_{y}} = -\beta\,( {y}_{o} -  {y}_{i})-(\alpha-\beta)\,  {e}_{iy} \,  \,[\,  {e}_{ix} \,  \, ( {x}_{o} -  {x}_{i})  \\+   {e}_{iy} \, \, ( {y}_{o} -  {y}_{i}) ] 
    \label{eqn:8D}
\end{multline}
\end{center}

\begin{figure}[H]%Unit Cell of Square Lattice
        \centering
        \includegraphics[height=6.0cm,width=5.86cm]{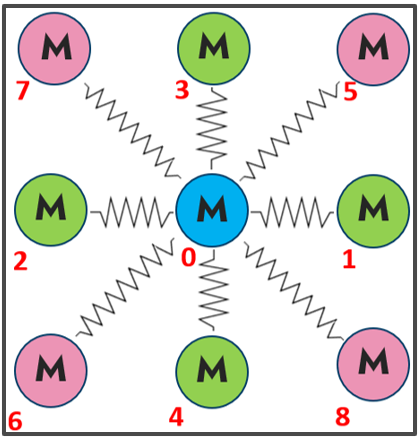}
        \caption{Unit cell of a square lattice showing the nearest neighbors $1,\ 2,\ 3,\ 4$ and next-nearest neighbors $5,\ 6,\ 7,\ 8$ with respect to atom $0$.}
        \label{fig:unit cell of a Square lattice}
    \end{figure}

\begin{figure}[H]%Repetition of Square Lattice
        \centering
        \includegraphics[height=7.0cm,width=6.86cm]{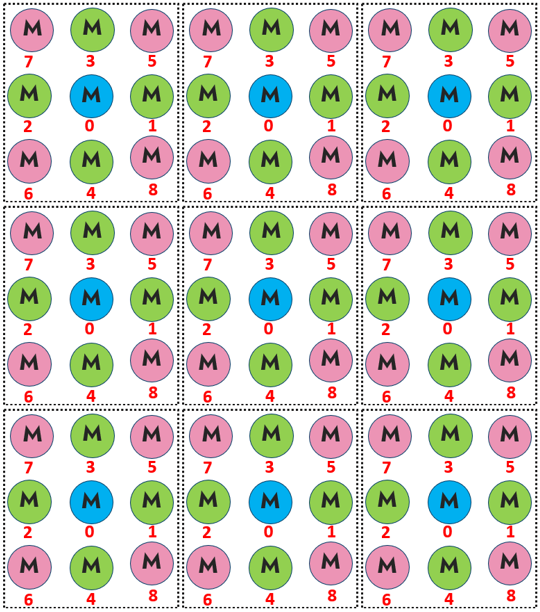}
        \caption{Construction of Square Lattice by repetition of unit cells: Each unit cell contains nine atoms indexed $0-8$.}
        \label{fig:Square Lattice Repetition}
    \end{figure}

To construct the computational model of the square lattice, we arbitrarily choose one atom as our origin (indexed as $0$) and assign indices ($1,\ 2,\ 3,\ 4$) to its nearest neighbor atoms and indices ($5,\ 6,\ 7,\ 8$) to the second nearest neighbor atoms as shown in Fig.~\ref{fig:unit cell of a Square lattice}. The unit cell thus contains nine atoms, each one vibrating with two degrees of freedom (DOF) along the two orthogonal (x \& y) axes. To span the planar infinite lattice, the unit cell is repeated in all the four directions as shown in Fig.~\ref{fig:Square Lattice Repetition}. The occupation number of the given unit cell is $4$ (one origin atom plus four corner atoms, each shared by four unit cells plus 4 edge atoms, each shared by two unit cells) with each atom having two DOF and so a monatomic square lattice is expected to exhibit $8$ phonon modes. 

In the harmonic approximation used for the model, the interactions between the atom with its nearest neighbors and next-nearest neighbors are modeled by elastic springs. The central force constants for nearest and next-nearest neighbor interactions are represented by $\alpha_1$, $\alpha_2$ respectively and the angular force constants by $\beta_1$, $\beta_2$ respectively for nearest and next-nearest neighbor interactions. The relative magnitudes of force constants between the nearest and next-nearest neighbors need to be commensurate with the corresponding interatomic separations. The nearest neighbors tend to have a stronger bonding as compared to next-nearest neighbors. In a real world situation, the exact ratio of the these force constants depends on many factors, such as, the physical nature of interatomic interactions, the specific atomic composition, the specific lattice structure. In the absence of a universal choice, we have used the ratio reported in the work by Cserti\cite{Cserti}. The nearest neighbor interactions are assumed to be twice as strong as the next-nearest neighbor interactions, so we have used $\alpha_1:\alpha_2$\,=\,$2:1$ and $\beta_1:\beta_2$\,=\,$2:1$ in our computational model.
 
Table \ref{table:Square lattice details} summarises the details related to the lattice dynamics of square lattice with respect to the arbitrarily chosen origin atom with index $0$ . The nearest neighbor distance is represented by $a$ and so the second nearest neighbor distance becomes $\sqrt{2}\ a$. The spatial coordinates of the neighboring atoms, the force constants involved and the direction cosines (DCs) of the equilibrium line joining the origin atom and its eight neighbors are listed in the table.
 \begin{table}[h!] %FINAL TAKEN
    \centering
    \renewcommand{\arraystretch}{1.2} % Adjust vertical spacing
      \begin{tabular}{|>{\centering\arraybackslash}m{0.7cm}|>{\centering\arraybackslash}m{1cm}|>{\centering\arraybackslash}m{1cm}|>{\centering\arraybackslash}m{1cm}|>{\centering\arraybackslash}m{1cm}|>{\centering\arraybackslash}m{1cm}|>{\centering\arraybackslash}m{1.1cm}|}
            \hline
            Index & \multicolumn{2}{|c|}{Force constants} & \multicolumn{2}{|c|}{Spatial Positions} & \multicolumn{2}{|c|}{DCs} \\ [10pt]
            \hline
             $i$ & Central & Angular &$p_{i}$ &$q_{i}$ &$e_{ix}$ & $e_{iy}$\\[10pt]
            \hline
            1 & $\alpha_{1}$ & $\beta_{1}$ & $a$ & $0$ & $1$ & $0$\\[8pt]
            \hline
            2 & $\alpha_{1}$ & $\beta_{1}$ & $-a$ & $0$ & $-1$ & $0$ \\[8pt]
            \hline
            3 & $\alpha_{1}$ & $\beta_{1}$ & $0$ &$a$ & $0$ & $1$ \\[8pt]
            \hline
            4 & $\alpha_{1}$ & $\beta_{1}$ & $0$ & $-a$ & $0$ & $-1$ \\[8pt]
            \hline
            5 & $\alpha_{2}$ & $\beta_{2}$ & $a$& $a$& $\frac{1}{\sqrt{2}}$&$\frac{1}{\sqrt{2}} $\\[8pt]
            \hline
            6 & $\alpha_{2}$ & $\beta_{2}$ & $-a$ & $-a$ &$-\frac{1}{\sqrt{2}}$&$-\frac{1}{\sqrt{2}} $ \\[8pt]
            \hline
            7 & $\alpha_{2}$ & $\beta_{2}$ & $-a$ & $a$ & $-\frac{1}{\sqrt{2}}$&$\frac{1}{\sqrt{2}} $ \\[8pt]
            \hline
            8 & $\alpha_{2}$ & $\beta_{2}$ & $a$ & $-a$ & $\frac{1}{\sqrt{2}}$&$-\frac{1}{\sqrt{2}} $ \\[8pt]
            \hline
        \end{tabular}
        \caption{Table listing the details related to the lattice dynamics of square lattice with respect to the origin atom with index $0$}
        \label{table:Square lattice details}
    \end{table}

 Using the expressions given in Eq.(\ref{eqn:8C}) and Eq.(\ref{eqn:8D}) for \textbf{\emph{x}} and \textbf{\emph{y}} components of the net force experienced by the atom $\bm{0}$ due to the atom $\bm{i}$ and the lattice details in table \ref{table:Square lattice details}, the Cartesian components of equations of motion for our reference atom $\bm{0}$ are formulated as follows:
\begin{multline}
        M \, \frac{{d^2x_{0}}}{{dt^2}} = - [\alpha_1(2x_0-x_1-x_2)+\beta_1(2x_0-x_3-x_4)
    \\+\frac{(\alpha_2+\beta_2)}{2}\,(4x_0-x_5-x_6-x_7-x_8)+\\+\frac{(\alpha_2-\beta_2)}{2}\,(-y_5-y_6+y_7+y_8)]
    \label{eqn:9}
\end{multline}
\begin{multline}
     M \, \frac{{d^2y_{0}}}{{dt^2}}= -[\alpha_1(2y_0-y_3-y_4)+\beta_1(2y_0-y_1-y_2)+\\+\frac{(\alpha_2+\beta_2)}{2}\,(4y_0-y_5-y_6-y_7-y_8)+\\+\frac{(\alpha_2-\beta_2)}{2}\,(-x_5-x_6+x_7+x_8)]
 \label{eqn:10}
\end{multline} 

\begin{table}[h!]
    \centering
    \renewcommand{\arraystretch}{1.2} % Adjust vertical spacing
      \begin{tabular}{|>{\centering\arraybackslash}m{2 cm}|>{\centering\arraybackslash}m{2.5 cm}|>{\centering\arraybackslash}m{3 cm}|}
          \hline
             Reference atom & Nearest neighbors & Next - nearest \linebreak neighbors  \\[12pt]
          \hline
           0 & 1,2,3,4 & 5,6,7,8\\[10pt]
          \hline
           1 & 0,2,5,8 & 3,4,6,7 \\[10pt]
          \hline
          2 & 0,1,6,7 & 3,4,5,8 \\[10pt]
          \hline
          3 & 0,4,5,7 & 1,2,6,8 \\[10pt]
          \hline
          4 & 0,3,6,8 & 1,2,5,7 \\[10pt]
          \hline
          5 & 1,3,7,8 & 0,2,4,6 \\[10pt]
          \hline
           6 & 2,4,7,8 & 0,1,3,5 \\[10pt]
          \hline
           7 & 2,3,5,6 & 0,1,4,8 \\[10pt]
          \hline
           8 & 1,4,5,6 & 0,2,3,7 \\[10pt]
          \hline
          
      \end{tabular}
    \caption{Table listing the nearest and the next-nearest neighbors of atoms in the unit cell of monatomic square lattice.}
    \label{table:Reference atom and neighbors}
  \end{table}
Implementation of Bvk periodic boundary conditions for the given lattice requires us to identify the neighbors of the unit cell atoms indexed $1-8$. Table \ref{table:Reference atom and neighbors} lists the nearest and next-nearest neighbors of these atoms using the periodicity of the lattice depicted in Fig.~\ref{fig:Square Lattice Repetition}. The equations of motion for the eight atoms can hence be constructed as done for the reference atom $\bm{0}$ in Eq.(\ref{eqn:9}) and Eq.(\ref{eqn:10}). Thus, we have a system of 18 coupled second order differential equations to be solved simultaneously in the displacement-time domain using the fourth order Runge Kutta method.
    
\begin{figure}[H]%Brillouin zone of Square Lattice
        \centering
        \includegraphics[height=5.6cm,width=6.6cm]{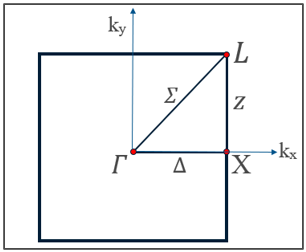}
        \caption{Brillouin zone of Square lattice: Annotated symmetry points are $\,\Gamma(0, 0),\, X\left(\frac{\pi}{a}, 0\right), \,L\left(\frac{\pi}{a}, \frac{\pi}{a}\right)$, $\triangle\,(\frac{\pi}{2a}, 0)$ and $\Sigma\,(\frac{\pi}{2a}, \frac{\pi}{2a})$}
        \label{fig:Brillouin zone of a Square lattice}
    \end{figure}

To aid the understanding of expected phonon spectrum, we depict the first Brillouin zone for the given monatomic square lattice with the high symmetry points in Fig.~\ref{fig:Brillouin zone of a Square lattice}. The ($k_x,k_y$) coordinates of the annotated symmetry points are $\,\Gamma(0, 0),\, X\left(\frac{\pi}{a}, 0\right), \,L\left(\frac{\pi}{a}, \frac{\pi}{a}\right)$, $\triangle\,(\frac{\pi}{2a}, 0)$ and $\Sigma\,(\frac{\pi}{2a}, \frac{\pi}{2a})$.

The traditional treatment of the problem by assuming plane wave solutions of the form in Eq.(\ref{eqn:0B}) for Eq.(\ref{eqn:9}) - Eq.(\ref{eqn:10}) has been done in the book authored by H.C.Gupta\cite{HC_Gupta}. The expression for secular determinant derived therein is given below in Eq.(\ref{eqn:secular determinant}):
\begin{center}
    \begin{equation}
        \begin{vmatrix}
                2\alpha\,(1-C_1) + 2\beta\,(1-C_2) & 2\,S_1\, \,S_2 (\alpha-\beta)\\
                + 2\,(\alpha+\beta)(1-C_1 \,C_2) & \\
                   - M\,\omega^2  & \\
               &  2\alpha\,(1-C_1) + 2\beta\,(1-C_2) \\
                2\,S_1\, \,S_2 (\alpha-\beta)  & + 2\,(\alpha+\beta)(1-C_1 \,C_2)\\
               & - M\,\omega^2
            \label{eqn:secular determinant}
        \end{vmatrix} = 0
    \end{equation}   
\end{center}
where the notations mean: $C_1=cos(a k_x)$, $C_2=cos(a k_y)$, $S_1=sin(a k_x)$, $S_2=sin(a k_y)$; ($k_x,\ k_y$) are the position coordinates of a point in the Brillouin zone. \\
\indent The expressions for the phonon frequencies corresponding to the high symmetry points in the FBZ are obtained by solving the determinant Eq.(\ref{eqn:secular determinant}) using the ($k_x,\ k_y$) coordinates for these points. The derived expressions are given in Table \ref{table:Phonon frequencies for monatomic square lattice with both types of forces}. The degeneracy of the longitudinal and transverse acoustical branches at the $\bm{L}$ point is to be noted. 
\begin{figure}[H]%Dispersion relation of monatomic Square Lattice
        \centering
        \includegraphics[height=5.0cm,width=7.159cm]{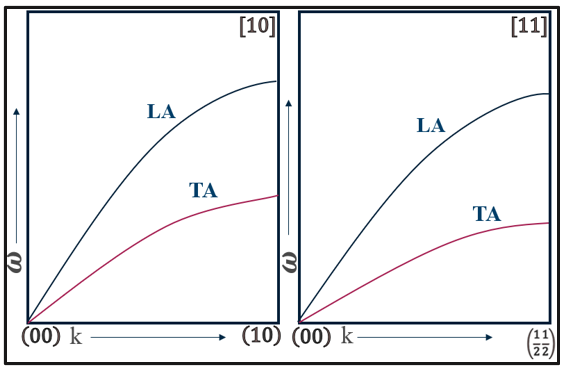}
        \caption{Phonon dispersion spectrum for a monatomic square lattice along the symmetry directions $[10]$ and $[11]$.}
        \label{fig:phonon dispersion in monatomic square lattice}
    \end{figure}  
The dispersion curves in the symmetry directions [1 0] and [1 1] are shown in Fig.~\ref{fig:phonon dispersion in monatomic square lattice}. Both the directions have one longitudinal acoustic (LA) and one transverse acoustic (TA) branch.
\begin{table}[h!]
    \centering
    \renewcommand{\arraystretch}{1.2} % Adjust vertical spacing
    \begin{tabular}{|>{\centering\arraybackslash}m{1.6 cm}|>{\centering\arraybackslash}m{1.6 cm}|>{\centering\arraybackslash}m{0.9 cm}|>{\centering\arraybackslash}m{3.5 cm}|}
      \hline
      Phonon Frequency & Symmetry Points & Branch type & Frequency \linebreak Expression \\[12pt]
      \hline
      $f_1^{L}$ & \multirow{2}{*}{\large $L\left(\frac{\pi}{a}, \frac{\pi}{a}\right)$} & LA   & $\displaystyle\frac{1}{2\pi} \sqrt{\frac{4\alpha_1+4\alpha_2}{m}}$ \\[  14pt]
      $f_2^{L}$ &&TA   & $\displaystyle\frac{1}{2\pi} \sqrt{\frac{4\alpha_1+4\alpha_2}{m}} $\\[  14pt]
      \hline
      $f_1^{X}$ & \multirow{2}{*}{\large $X\left(\frac{\pi}{a}, 0\right)$} & LA   & $\displaystyle\frac{1}{2\pi} \sqrt{\frac{4\alpha_1+4\beta_1+4\beta_2}{m}}$ \\[  14pt]
      $f_2^{X}$ &&TA   & $\displaystyle\frac{1}{2\pi} \sqrt{\frac{4\alpha_2+4\beta_1+4\beta_2}{m}} $\\[  14pt]
      \hline
       $f_1^{\triangle}$ & \multirow{2}{*}{\large$\triangle\,(\frac{\pi}{2a}, 0)$} & LA   & $\displaystyle\frac{1}{2\pi} \sqrt{\frac{2\alpha_1+2\beta_1+2\beta_2}{m}}$ \\[  14pt]
      $f_2^{\triangle}$ &&TA   & $\displaystyle\frac{1}{2\pi} \sqrt{\frac{2\alpha_2+2\beta_1+2\beta_2}{m}} $\\[  14pt]
      \hline
      $f_1^{\Sigma}$ & \multirow{2}{*}{\large$\Sigma\,(\frac{\pi}{2a}, \frac{\pi}{2a})$} & LA   & $\displaystyle\frac{1}{2\pi} \sqrt{\frac{2\alpha_1+2\beta_1+4\alpha_2}{m}}$ \\[  14pt]
      $f_2^{\Sigma}$ &&TA   & $\displaystyle\frac{1}{2\pi} \sqrt{\frac{2\alpha_1+2\beta_1+4\beta_2}{m}}$\\[  14pt]
      \hline
    \end{tabular}
    \caption{Table listing the analytical expressions for phonon frequencies at the high symmetry points in a monatomic square lattice in terms of nearest and next-nearest neighbor central force and angular force constants $\alpha_1,\ \alpha_2, \beta_1 \ \& \beta_2$ respectively.}
    \label{table:Phonon frequencies for monatomic square lattice with both types of forces}
\end{table}

\begin{figure}[H]%2D Square Lattice FFT with only central
        \centering
        \includegraphics[height=7.4cm,width=8.6cm]{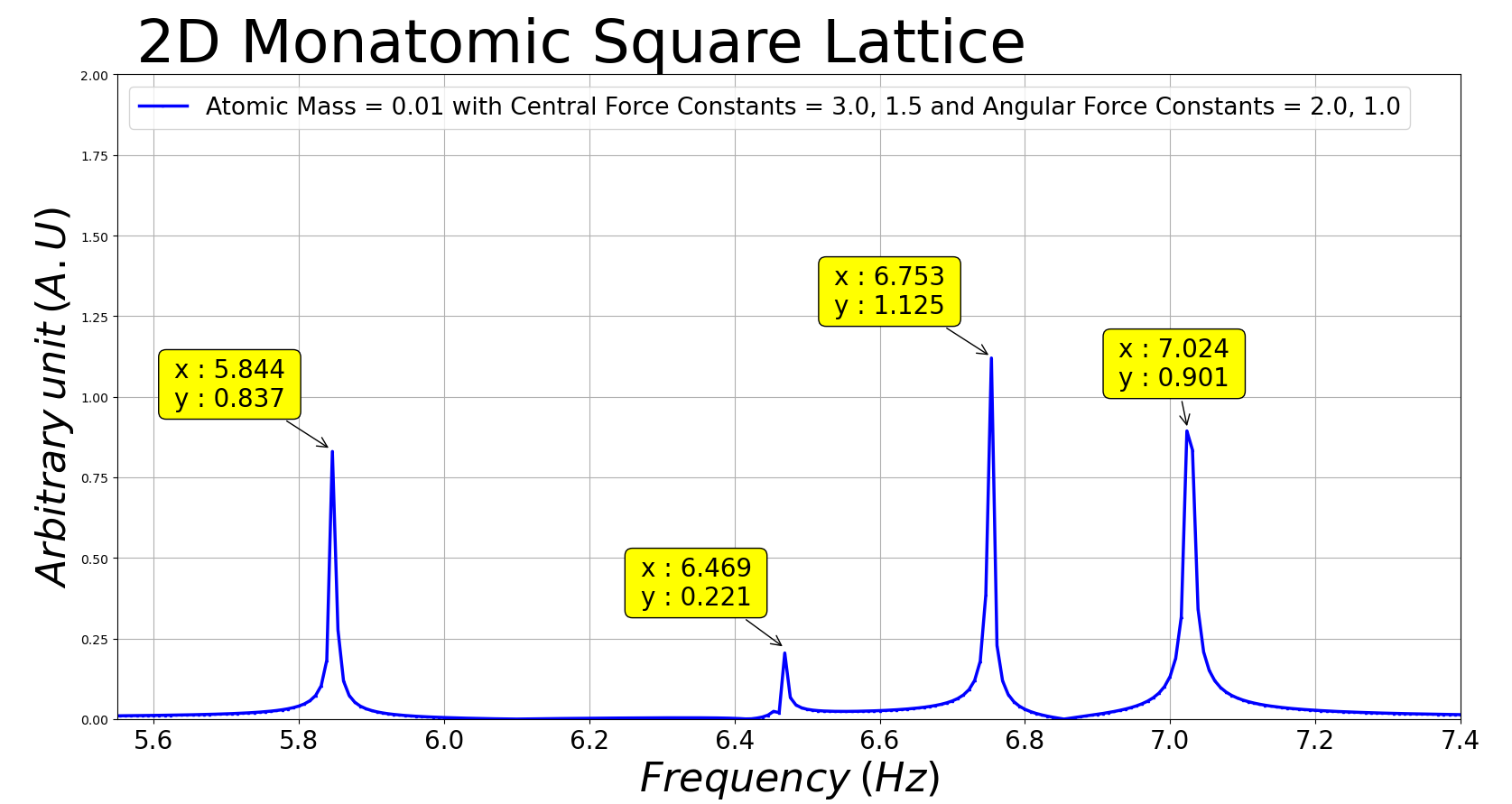}
        \caption{FFT plot for an atom of a monatomic square lattice modeled using central and angular forces: Each atom in the unit cell exhibits $4$ peaks.}
        \label{fig:FFT plot for monatomic square lattice}
    \end{figure}

The FFT computation of the instantaneous displacement solutions obtained for the system of $18$ coupled equations of motion for the $9$ atoms in the unit cell exhibit $4$ peaks for each atom, irrespective of the randomized set of initial conditions. The plot for each atom exhibits FFT peaks of varying heights but at the same frequency values. As discussed earlier, the relative heights of FFT peaks are only indicative of the degree of participation of a given atom in a given normal (phonon) mode of vibration. This aspect of the problem is insignificant in the context of the present work. Hence, we depict the FFT plot of only one of the atoms in Fig.~\ref{fig:FFT plot for monatomic square lattice}. The values of the model parameters atomic mass and force constants used for computation are: $m = 0.01  \,\,; \,\, \alpha_1 = 3.0 \,\,;\,\, \alpha_2 = 1.5 \,\, ;\,\, \beta_1 = 2.0 \,\,; \,\, \beta_2 = 1.0 $.
%%%%%%%%%%%%%%%%%%%%%%%%%%%%%%%%%%%%%%%%%%%%%%%%%%%%%%%%%%%%%%%%%%%%%%

A mapping and comparison of computed phonon frequencies with the values calculated using the analytical frequency expressions in Table \ref{table:FFT Table for monatomic square lattice}. It is seen that our FFT computation exactly captures the degenerate \emph{LA} and \emph{TA} branch frequency at the \emph{L}-point. The phonon frequencies at the \emph{$\Sigma$}-point are also captured quite accurately within an absolute error of $2\%$. For the given choice of force constant ratios used in the computation, the \emph{TA} branch at \emph{X} point becomes degenerate with the \emph{L}-point \emph{LA}, \emph{TA} branches and so is also captured exactly.

\begin{table}[h!]
    \centering
    \renewcommand{\arraystretch}{1.2} % Adjust vertical spacing
    \begin{tabular}{|>{\centering\arraybackslash}m{2 cm}|>{\centering\arraybackslash}m{2 cm}|>{\centering\arraybackslash}m{2 cm}|>{\centering\arraybackslash}m{2 cm}|}
      \hline
      \multicolumn{4}{|c|}{Theoretical Phonon Frequencies at the Symmetry Points} \\[13pt]
      \hline
      \large $L\left(\frac{\pi}{a}, \frac{\pi}{a}\right)$ & \large $X\left(\frac{\pi}{a}, 0\right)$ & \large$\triangle\,(\frac{\pi}{2a}, 0)$ & \large$\Sigma\,(\frac{\pi}{2a}, \frac{\pi}{2a})$ \\[ 12pt]
       \hline
       $f_1^{L} = 6.752$ &  $f_1^{X} = 7.797$ &  $f_1^{\triangle} = 5.513$ & $f_1^{\Sigma} = 6.366$ \\[ 10pt]
       $f_2^{L} = 6.752$ &  $f_2^{X} = 6.752$ &  $f_2^{\triangle} = 4.775$ & $f_2^{\Sigma} = 5.955$ \\[ 10pt]
       \hline
       \multicolumn{4}{|c|}{Phonon Frequencies captured by FFT} \\[13pt]
       \hline
       $f_1^{L'} =  \textcolor{red}{6.753} \linebreak (\textcolor{blue}{0.01\%}) $ &  $f_1^{X'} = \textcolor{red}{7.024} \linebreak (\textcolor{blue}{9.90\%})$ &  $ - $ & $f_1^{\Sigma'} =  \textcolor{red}{6.469} \linebreak (\textcolor{blue}{1.62\%})$ \\[ 10pt]
       $f_2^{L'} = \textcolor{red}{6.753}  \linebreak (\textcolor{blue}{0.01\%})$ &  $f_2^{X'} = \textcolor{red}{6.753}  \linebreak (\textcolor{blue}{0.01\%})$ &  $ - $ & $f_2^{\Sigma'} = \textcolor{red}{5.844} \linebreak (\textcolor{blue}{1.86\%})$ \\[ 10pt]
       \hline
    \end{tabular}
    \caption{Table lists the expected theoretical phonon frequency values and corresponding computational values for a monatomic square lattice when considering both central and angular force interactions (The percentages within brackets denote the absolute percentage errors).}
    \label{table:FFT Table for monatomic square lattice}
\end{table}

The limited existing literature\cite{Cserti}$^{,}$\cite{papoular}$^{,}$\cite{Iachello} available on monatomic square lattice models employ only the central forces to account for nearest and next-nearest neighbor interatomic interactions. Our model can be easily reduced to the special case of central force approximation for interatomic interactions up to the second nearest neighbors. This is achieved by putting $\beta_1=\beta_2=0$ in Eq.(\ref{eqn:9}) - Eq.(\ref{eqn:10}) and the remaining $16$ equations of motion for atoms indexed $1-8$. The corresponding analytical expressions for the $8$ phonon frequencies at the FBZ symmetry points are given in Table \ref{table:FFT Table for monatomic square lattice with only central force}. The \emph{LA} and \emph{TA} branches at the \emph{L}-point again seen to be degenerate. Further, for our given choice of the ratio of two central force constants ($\alpha_1=2\alpha_2$), the degeneracy is reflected in other phonon modes too. The \emph{TA} branch frequencies at \emph{X} and \emph{$\Sigma$} points become equal to the \emph{LA} branch frequency at \emph{$\Delta$}-point . Similarly, the \emph{LA} branches at \emph{X} and \emph{$\Sigma$} points become degenerate.
\begin{table}[h!]
    \centering
    \renewcommand{\arraystretch}{1.2} % Adjust vertical spacing
    \begin{tabular}{|>{\centering\arraybackslash}m{1.6 cm}|>{\centering\arraybackslash}m{1.6 cm}|>{\centering\arraybackslash}m{1 cm}|>{\centering\arraybackslash}m{3.5 cm}|}
      \hline
      Phonon Frequency & Symmetry Points & Branch type & Frequency Expression \\[12pt]
      \hline
      $f_1^{L}$ & \multirow{2}{*}{\large $L\left(\frac{\pi}{a}, \frac{\pi}{a}\right)$} & LA   & $\displaystyle\frac{1}{2\pi} \sqrt{\frac{4\alpha_1+4\alpha_2}{m}}$ \\[  10pt]
      $f_2^{L}$ &&TA   & $\displaystyle\frac{1}{2\pi} \sqrt{\frac{4\alpha_1+4\alpha_2}{m}} $\\[  10pt]
      \hline
      $f_1^{X}$ & \multirow{2}{*}{\large $X\left(\frac{\pi}{a}, 0\right)$} & LA   & $\displaystyle\frac{1}{2\pi} \sqrt{\frac{4\alpha_1}{m}}$ \\[  10pt]
      $f_2^{X}$ &&TA   & $\displaystyle\frac{1}{2\pi} \sqrt{\frac{4\alpha_2}{m}} $\\[  10pt]
      \hline
       $f_1^{\triangle}$ & \multirow{2}{*}{\large$\triangle\,(\frac{\pi}{2a}, 0)$} & LA   & $\displaystyle\frac{1}{2\pi} \sqrt{\frac{2\alpha_1}{m}}$ \\[  10pt]
      $f_2^{\triangle}$ &&TA   & $\displaystyle\frac{1}{2\pi} \sqrt{\frac{2\alpha_2}{m}} $\\[  10pt]
      \hline
      $f_1^{\Sigma}$ & \multirow{2}{*}{\large$\Sigma\,(\frac{\pi}{2a}, \frac{\pi}{2a})$} & LA   & $\displaystyle\frac{1}{2\pi} \sqrt{\frac{2\alpha_1+4\alpha_2}{m}}$ \\[  10pt]
      $f_2^{\Sigma}$ &&TA   & $\displaystyle\frac{1}{2\pi} \sqrt{\frac{2\alpha_1}{m}}$\\[  10pt]
      \hline
    \end{tabular}
    \caption{Table listing the analytical expressions for phonon frequencies at the high symmetry points in a monatomic square lattice in terms of nearest neighbor and next-nearest neighbor central force constants $\alpha_1$ and $\alpha_2$ respectively.}
    \label{table:Phonon frequencies for monatomic square lattice with only central}
\end{table}
\begin{figure}[H]%2D 
        \centering
        \includegraphics[height=6.5cm,width=8.1cm]{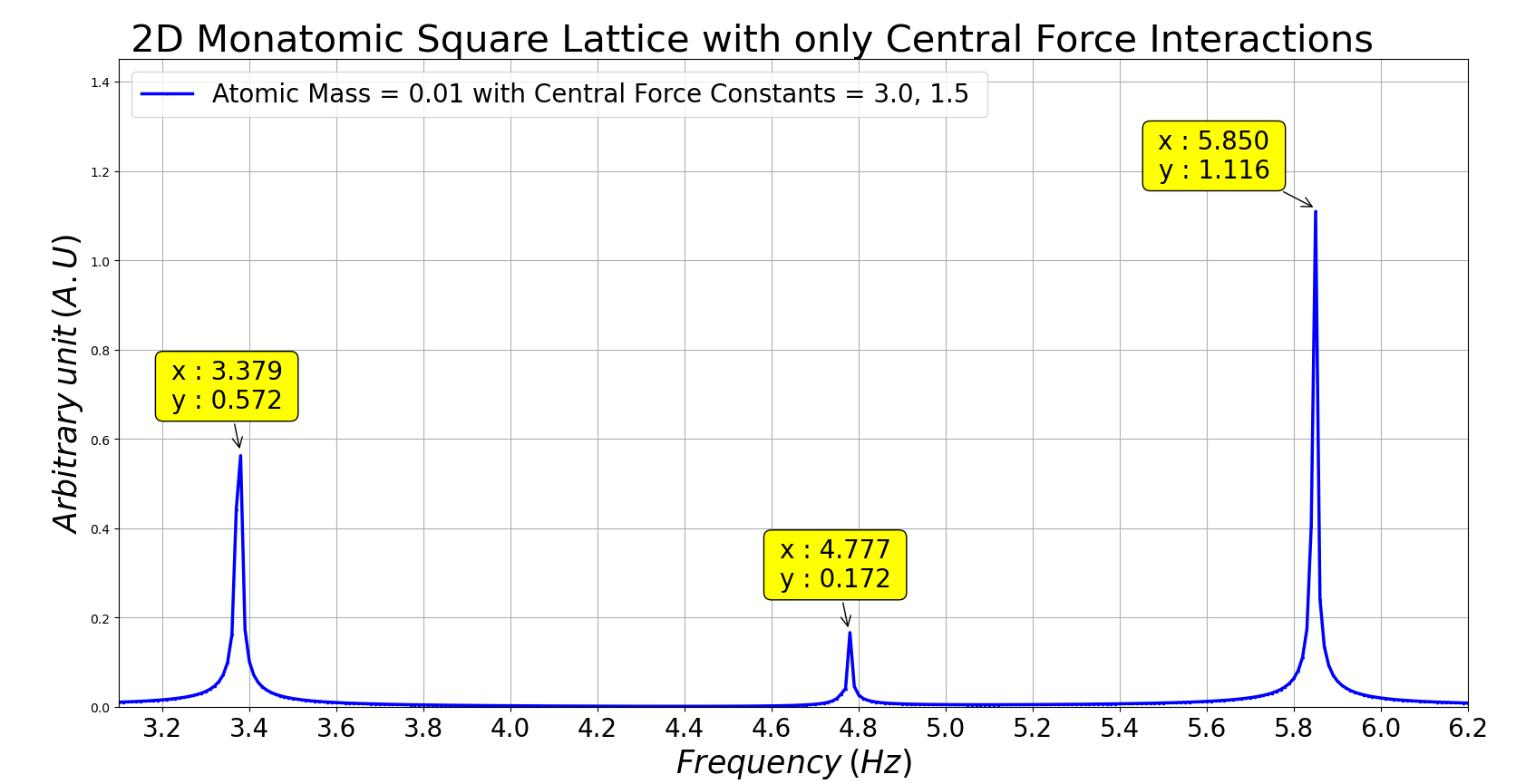}
        \caption{FFT plot for an atom of a monatomic square lattice modeled using only central forces: Each atom in the unit cell exhibits $3$ peaks.}
        \label{fig:FFT plot for monatomic square lattice with only central}
    \end{figure}
The FFT computation using the model parameter values $m=0.01$, $\alpha_1=3.0$, $\alpha_2=1.5$ yield $3$ distinct peaks for each of the $9$ atoms in the unit cell. Figure~\ref{fig:FFT plot for monatomic square lattice with only central} depicts the FFT plot for one of the atoms in the unit cell of the given monatomic square lattice. The mapping and comparison of the computed and analytical phonon frequencies is given in table \ref{table:FFT Table for monatomic square lattice with only central force}. All the FFT computed peaks can be seen to exhibit relatively higher absolute errors of around $13\%$. 

It is evident from the results summarized in Tables \ref{table:FFT Table for monatomic square lattice} and \ref{table:FFT Table for monatomic square lattice with only central force} that the phonon dynamics of a monatomic square lattice is better explained by modeling the interatomic interactions in terms of central and angular forces. The incorporation of angular forces lifts the degeneracy of the phonon modes that is encountered in the model using only the central forces for nearest and next-nearest neighbor interactions. However, in both the models of monatomic square lattice discussed, the FFT spectrum has some missing values as indicated by dashes in tables \ref{table:FFT Table for monatomic square lattice} and \ref{table:FFT Table for monatomic square lattice with only central force}. The failure of FFT algorithm to capture the missing phonon frequencies is attributed to the limitation of the lattice dynamical model to account for interatomic interactions responsible for the missing phonon modes.  

\begin{table}[h!]
    \centering
    \renewcommand{\arraystretch}{1.2} % Adjust vertical spacing
    \begin{tabular}{|>{\centering\arraybackslash}m{2 cm}|>{\centering\arraybackslash}m{2 cm}|>{\centering\arraybackslash}m{2 cm}|>{\centering\arraybackslash}m{2 cm}|}
      \hline
      \multicolumn{4}{|c|}{Theoretical Phonon Frequencies at the Symmetry Points} \\[13pt]
      \hline
      \large $L\left(\frac{\pi}{a}, \frac{\pi}{a}\right)$ & \large $X\left(\frac{\pi}{a}, 0\right)$ & \large$\triangle\,(\frac{\pi}{2a}, 0)$ & \large$\Sigma\,(\frac{\pi}{2a}, \frac{\pi}{2a})$ \\[ 12pt]
       \hline
       $f_1^{L} = 6.752$ &  $f_1^{X} = 5.513$ &  $f_1^{\triangle} = 3.898$ & $f_1^{\Sigma} = 5.513$ \\[ 10pt]
       $f_2^{L} = 6.752$ &  $f_2^{X} = 3.898$ &  $f_2^{\triangle} = 2.757$ & $f_2^{\Sigma} = 3.898$ \\[ 10pt]
       \hline
       \multicolumn{4}{|c|}{Phonon Frequencies captured by FFT} \\[13pt]
       \hline
       $f_1^{L'} = \textcolor{red}{5.850} \linebreak (\textcolor{blue}{13.36\%}) $ &  $ f_1^{X'} = \textcolor{red}{4.777} \linebreak (\textcolor{blue}{13.35\%})$  &  $ f_1^{\triangle'} = \textcolor{red}{3.379} \linebreak (\textcolor{blue}{13.31\%}) $ & $f_1^{\Sigma'} = \textcolor{red}{4.777}\linebreak (\textcolor{blue}{13.35\%})$ \\[ 10pt]
       $f_2^{L'} = \textcolor{red}{5.850} \linebreak (\textcolor{blue}{13.36\%})$ &  $f_2^{X'} = \textcolor{red}{3.379}\linebreak (\textcolor{blue}{13.31\%})$ &  $ - $ & $f_2^{\Sigma'}  = \textcolor{red}{3.379} \linebreak (\textcolor{blue}{13.31\%})$ \\[ 10pt]
       \hline
    \end{tabular}
    \caption{Table lists the expected theoretical phonon frequency values and corresponding computational values for a monatomic square lattice when considering only central force interactions (The percentages within brackets denote the absolute percentage errors).}
    \label{table:FFT Table for monatomic square lattice with only central force}
  \end{table}
%\linebreak (\textcolor{blue}{0.01\%})

\section{\label{sec6:}Monatomic Honeycomb Lattice} \protect

Unlike the monatomic square lattice which has more of a pedagogical importance in providing an in-depth understanding of the theory of lattice dynamics, monatomic honeycomb lattices have gathered considerable attention in the real world. Graphene is a single layer of carbon atoms arranged in a two-dimensional honeycomb lattice. The material exhibits exceptional tensile strength, flexibility and transparency, high electrical and thermal conductivity, fluid impermeability and alongside some unique optical properties. Indubitably, graphene is often referred to as a wonder material in literature. It forms the basic structural element of other carbon allotropes like graphite\cite{Cserti}$^{,}$\cite{lang}$^{,}$\cite{Maeda}$^{,}$\cite{Gupta_malhotra}, carbon nanotubes\cite{Woods}$^{,}$\cite{Damnjanovic}$^{,}$\cite{Jishi} and fullerenes\cite{Mousavi}$^{,}$\cite{Matija} that have potential industrial applications.
\begin{figure}[H]%2D Hexagonal lattice
        \includegraphics[height=4.04cm,width=8.1cm]{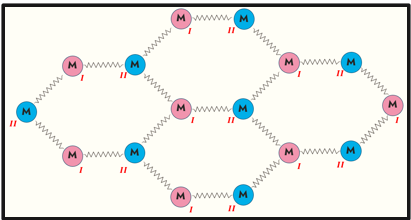}
        \caption{Hexagonal lattice is a non-Bravais lattice: Lattice sites labelled $I$ and $II$ are not equivalent due to the different orientation of adjacent neighbors at the respective sites.}
        \label{fig:2D Hexagon lattice}
    \end{figure}
In this section we extend our model for computing the phonon spectrum using BvK boundary conditions to a honeycomb lattice. It is a non-Bravais lattice comprising of lattice sites at the corners of hexagonal unit cells. Our model assumes that each of these lattice sites are occupied by identical atoms of mass $M$ each, held together by elastic springs as shown in Fig.~\ref{fig:2D Hexagon lattice}. It can be seen that the lattice sites labelled $I$ and $II$ are not equivalent due to the different orientation of adjacent neighbors at the respective sites. The dynamics of the lattice is calculated using central force type interactions between the nearest and the next-nearest neighbor atoms. The occupation number of the hexagonal unit cell is $2$ as each of the six corner atoms is shared by $3$ adjacent unit cells. Further, each of the two unit cell atoms has two DOF, so we expect the honeycomb lattice to exhibit $4$ phonon modes.  
 \begin{figure}[H]%2D Unit Cell Hexagonal lattice
        \includegraphics[height=4.84cm,width=8.43cm]{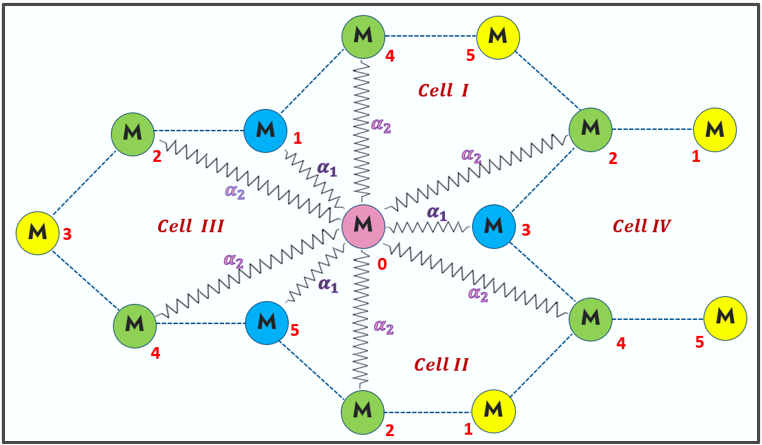}
        \caption{Unit cell of Honeycomb lattice with nearest and next-nearest neighbor interactions.}
        \label{fig:2D Unit Cell Hexagon lattice}
    \end{figure}
Figure~\ref{fig:2D Unit Cell Hexagon lattice} depicts three complete unit cells labelled $I$, $II$, $III$ and an incomplete cell labelled $IV$. Each atom sitting at a given lattice site has three nearest neighbors occupying its non-equivalent lattice sites  and six next-nearest neighbors at its equivalent lattice sites. For the atom indexed \emph{0}, the atoms with indices \emph{1,3,5} are the nearest neighbors connected by springs with force constant $\alpha_{1}$ and the atoms with indices \emph{2,4} connected by springs with force constant $\alpha_{2}$ are the next-nearest neighbors. The implementation of the BvK conditions is reflected by lattice sites with the same index number. If the given lattice had infinite spatial extent, then the atoms sitting at these lattice sites would be equivalent to each other and exhibit identical dynamics. The unit cell of our computational model thus contains six atoms indexed $0$-$5$. The nearest and next-nearest neighbors of each of these atoms in the unit cell are listed in Table\ref{table:Reference atom and neighbors for honeycomb}.
\begin{table}[h!]
    \centering
    \renewcommand{\arraystretch}{1.2} % Adjust vertical spacing
    \begin{tabular}{|>{\centering\arraybackslash}m{1.5cm}|>{\centering\arraybackslash}m{2.5cm}|>{\centering\arraybackslash}m{3cm}|}
          \hline
             Reference atom & Nearest \linebreak neighbors & Next - nearest neighbors\\[2pt]
             & & [Each index to represent 3 atoms]  \\[3pt]
          \hline
           0 & 1,3,5 & 2,4 \\[8pt]
          \hline
           1 & 0,2,4 & 3,5 \\[8pt]
          \hline
           2 & 1,3,5  & 0,4 \\[8pt]
          \hline
           3 & 0,2,4 & 1,5 \\[8pt]
          \hline
           4 & 1,3,5 & 0,2 \\[8pt]
          \hline
           5 & 0,2,4 & 1,3 \\[8pt]
          \hline
      \end{tabular}
    \caption{Table listing the nearest and the next-nearest neighbors of six atoms indexed in the unit cell of honeycomb lattice.}
    \label{table:Reference atom and neighbors for honeycomb}
  \end{table}

\begin{figure}[H]%FFT Hexagonal lattice
   \centering
        \includegraphics[height=5.1cm,width=8.4cm]{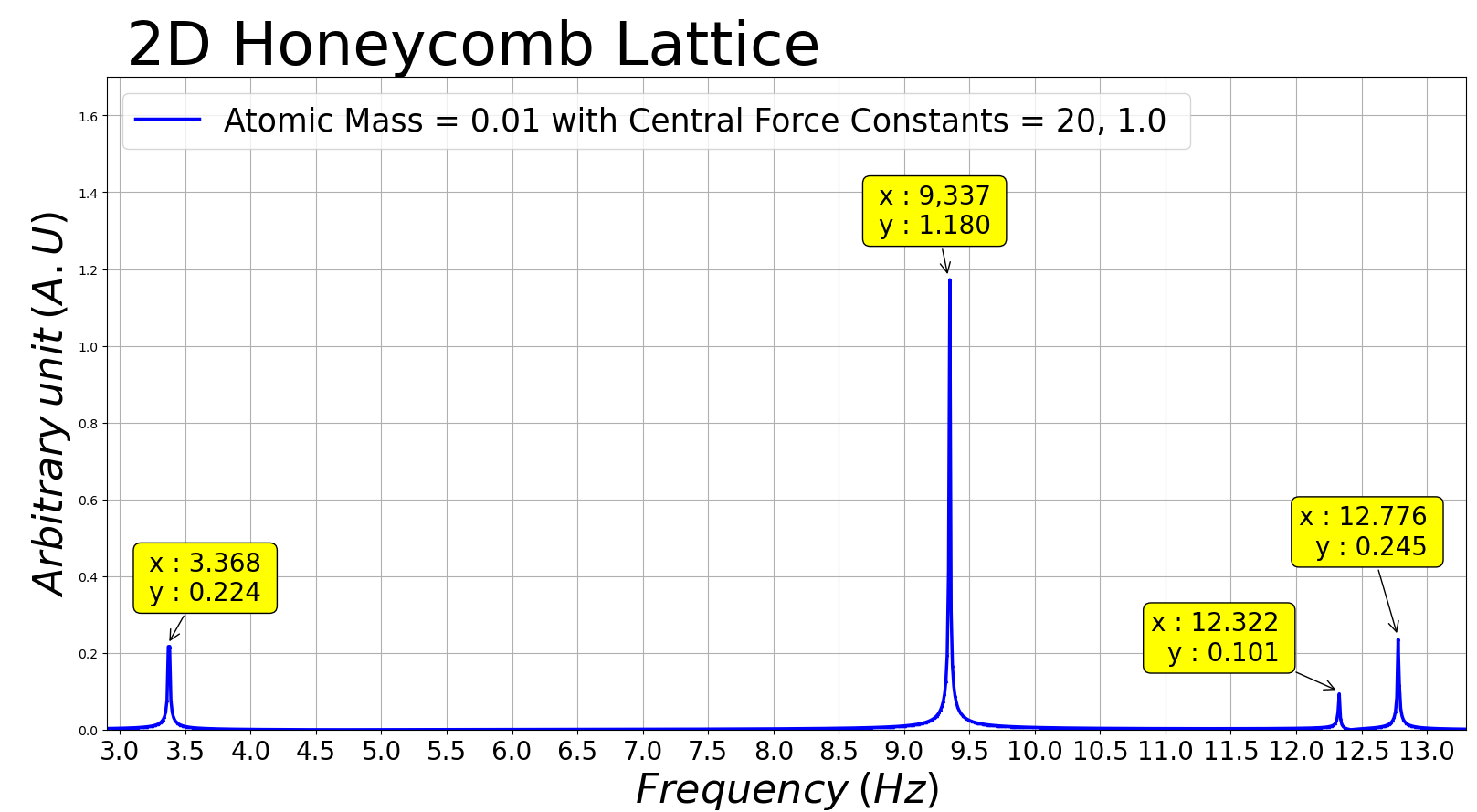}
        \caption{FFT plot of honeycomb lattice with force constant ratio 20:1.}
        \label{fig:FFT of Hexagon lattice a1=20a2}
    \end{figure}

Table \ref{table:Hexagonal lattice details} summarises the lattice dynamical details of the given honeycomb lattice with respect to the arbitrarily chosen origin atom with index \emph{0}. The nearest neighbor distance equal to the edge length of the hexagonal unit cell is represented by $a$, so the second nearest neighbor distance becomes $\sqrt{3}\ a$. The spatial coordinates of the neighboring atoms, the force constants and the direction cosines (DCs) of the equilibrium line joining the origin atom to its nine neighbors are listed in the table. 
\begin{table}[h!]
    \centering
    \renewcommand{\arraystretch}{1.2} % Adjust vertical spacing
    \begin{tabular}{|>{\centering\arraybackslash}m{0.7cm}|>{\centering\arraybackslash}m{1.4cm}|>{\centering\arraybackslash}m{1.2cm}|>{\centering\arraybackslash}m{0.8cm}|>{\centering\arraybackslash}m{0.8cm}|>{\centering\arraybackslash}m{0.7cm}|>{\centering\arraybackslash}m{0.7cm}|}
        \hline
        Index & Hexagonal Cells & Force constants & \multicolumn{2}{c|}{Spatial Positions} & \multicolumn{2}{c|}{DCs} \\
        \hline
        $i$ & Cell Index & Central Force  &$p_{i}$ &$q_{i}$ &$\hat{\mathbf{e}}_{xi}$ & $\hat{\mathbf{e}}_{yi}$\\
        \hline
        1 & $I, \,III$ & $\alpha_{1}$ & $-\frac{a}{2}$ & $\frac{\sqrt{3}\,a}{2}$ & $-\frac{1}{2}$ & $\frac{\sqrt{3}}{2}$\\
        \hline
        3 & $I, \,IV$ & $\alpha_{1}$ & $a$ & $0$ & $1$ & $0$ \\
        \hline
        5 & $II, \,III$ & $\alpha_{1}$ & $-\frac{a}{2}$ &$-\frac{\sqrt{3}\,a}{2}$ & $-\frac{1}{2}$ & $-\frac{\sqrt{3}}{2}$ \\
        \hline
        2 & $III$ & $\alpha_{2}$ & $-\frac{3a}{2}$& $\frac{\sqrt{3}\,a}{2}$& $-\frac{\sqrt{3}}{2}$&$\frac{1}{2}$\\
        \hline
        4 & $III$ & $\alpha_{2}$ & $-\frac{3a}{2}$ & $-\frac{\sqrt{3}\,a}{2}$ &$-\frac{\sqrt{3}}{2}$&$-\frac{1}{2} $ \\
        \hline
        2 & $I, \, IV$ & $\alpha_{2}$ & $\frac{3a}{2}$& $\frac{\sqrt{3}\,a}{2}$& $\frac{\sqrt{3}}{2}$&$\frac{1}{2}$\\
        \hline
        4 & $II, \,IV$ & $\alpha_{2}$ & $\frac{3a}{2}$ & $-\frac{\sqrt{3}\,a}{2}$ &$\frac{\sqrt{3}}{2}$&$-\frac{1}{2} $ \\
        \hline
        2 & $II$ & $\alpha_{2}$ & $0$ & $-{\sqrt{3}a}$ & $0$&$-1$ \\
        \hline
        4 & $I$ & $\alpha_{2}$ & $0$ & ${\sqrt{3}a}$ & $0$& $1$ \\
        \hline
    \end{tabular}
     \caption{Table listing the details related to the lattice dynamics of monatomic honeycomb lattice with respect to the origin atom with index $0$.}
     \label{table:Hexagonal lattice details}
    \end{table}
Using these details, the Cartesian components of equations of motion for our reference atom $\bm{0}$ are formulated as follows: \\
\noindent Equation of motion in \textbf{x} direction for atom $0$:
\begin{multline}
        m \, \frac{{d^2x_{0}}}{{dt^2}} = -[\left(\alpha_1\right)\left(x_0-x_3\right)+\left(\alpha_1\right)\left(\frac{\sqrt{3}}{4}\right)\left(y_1-y_5\right)+\\+\frac{\left(\alpha_1\right)}{4}\left(2x_0-x_1-x_5\right)+\left(\alpha_2\right)\left(\frac{3}{4}\right)\left(4x_0-2x_2-2x_4\right)]
        \label{eqn:20A}
\end{multline} 
\noindent Equation of motion in \textbf{y} direction for atom $0$:
\begin{multline}
         m \, \frac{{d^2y_{0}}}{{dt^2}}= -[\left(\alpha_1\right)\left(\frac{3}{4}\right)\left(2y_0-y_1-y_5\right)+\left(\alpha_2\right)\left(2y_0-y_2-y_4\right)+\\+\left(\alpha_1\right)\left(\frac{\sqrt{3}}{4}\right)\left(x_1- x_5\right)+\left(\alpha_2\right)\left(\frac{1}{4}\right)\left(4y_0-2y_2-2y_4\right)]
         \label{eqn:20B}
\end{multline}  
Similar equations of motion are formulated by choosing each of the remaining five atoms in the unit cell. Hence, we get a system of \emph{12} coupled second order differential equations which are solved simultaneously in displacement-time domain using the fourth order Runge-Kutta algorithm. The computation is done using $m=0.01$ for the atomic mass parameter and for different ratios of the force constants ($\alpha_1 : \alpha_2$).  
%%%%%%%%%%%%%%%%%%%%%%%%%%%%%%%%%%%%%%%%%%%%%%%%%%%%%%%%%%%%%%%%%%%%%%%
\begin{figure}[H]%FFT Hexagonal lattice
   \centering
        \includegraphics[height=5.1cm,width=8.4cm]{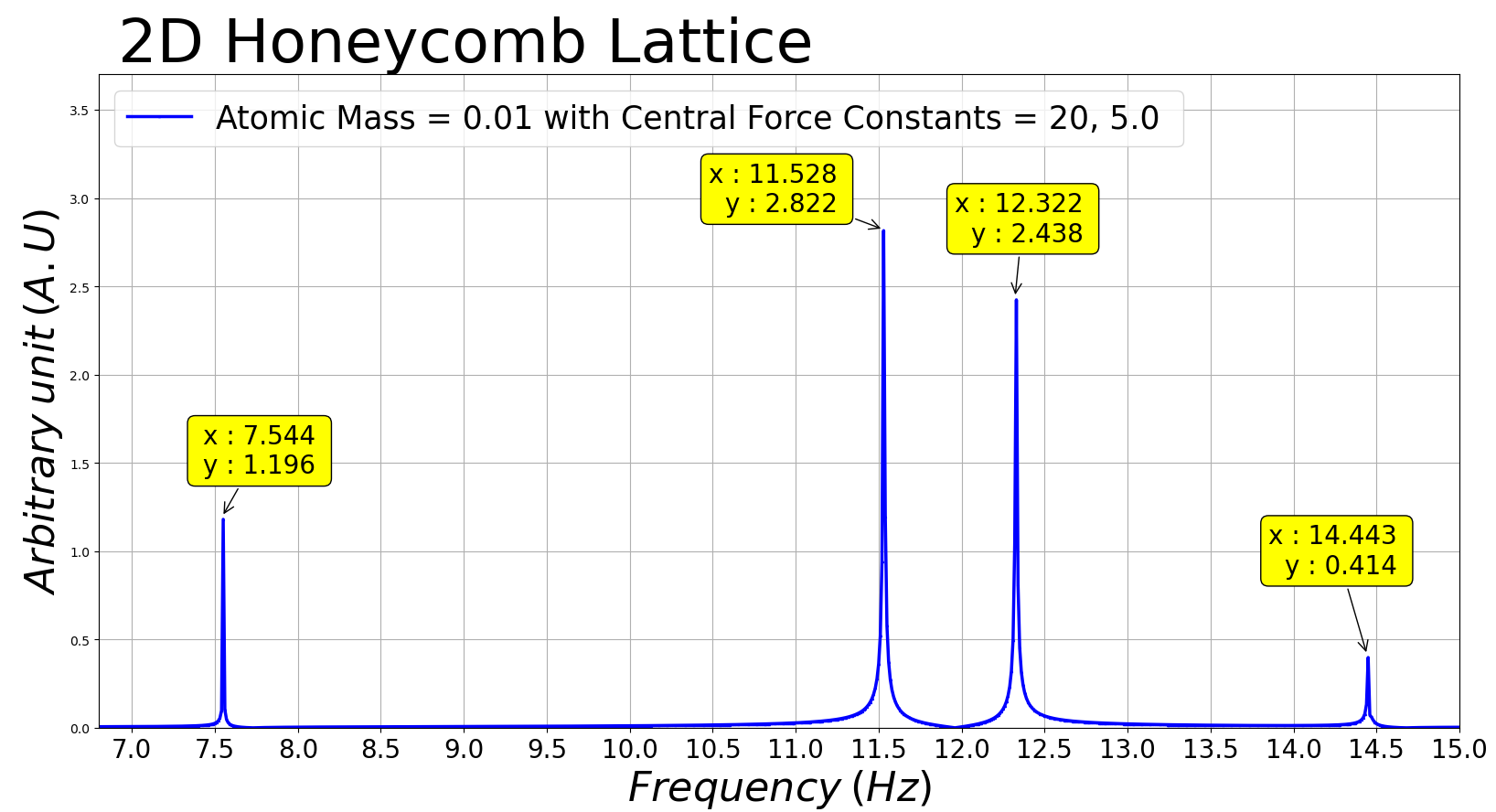}
        \caption{FFT plot of honeycomb lattice with force constant ratio 4:1.}
        \label{fig:FFT of Hexagon lattice a1=4a2}
    \end{figure}
The FFT computation of the instantaneous displacement solutions obtained for the system of \emph{12} coupled equations of motion for the six atoms in the unit cell exhibit $4$ peaks for each atom, irrespective of the randomized set of initial conditions and our choice of force constant ratio ($\alpha_1 : \alpha_2$). Once again, the relative heights of FFT peaks are only indicative of the degree of participation of a given atom in a given normal (phonon) mode of vibration and this aspect of the problem is insignificant in the context of the present work. Hence, we depict the FFT plots of only one of the unit cell atoms in Fig.~\ref{fig:FFT of Hexagon lattice a1=4a2} and Fig.~\ref{fig:FFT of Hexagon lattice a1=20a2} corresponding to the force constant ratios ($\alpha_1 : \alpha_2=4 : 1$) and ($\alpha_1 : \alpha_2=20 : 1$) respectively. The other ratio values that were explored in computation are ($\alpha_1 : \alpha_2=2:1,\ 6:1, \ 8:1,\ 10:1,\ \& 12:1$). The FFT computation of displacement-time solution in each case revealed only $4$ distinct peaks.

\begin{table}[h!]
    \centering
    \renewcommand{\arraystretch}{1.2} % Adjust vertical spacing
      \begin{tabular}{|>{\raggedright\arraybackslash}m{8 cm}|}
          \hline
             Elements of the Secular Determinant (S)\\[10pt]
          \hline
           $S_{11} =  S_{33} = 
           \displaystyle{M\,\omega^{2} - \frac{3}{2}\,\alpha_{1} - 3\,\alpha_2 + \frac{3}{2}\,\alpha_2\,(C_1+C_2)}$\\[10pt]
          \hline
           $S_{12} = S_{21} = \displaystyle{-\frac{\sqrt{3}}{2}\,\alpha_2\,(C_2-C_1)}$ \\[10pt]
          \hline
          $S_{13} = S_{31}^{*} = \displaystyle{\alpha_1\,e^{i k_x a}+\frac{\alpha_1}{2}\,e^{-i (\frac{k_x\,a}{2})}\,C_3}$ \\[10pt]
            \hline
          $S_{14} =  S_{41}^{*} = \displaystyle{-\frac{\sqrt{3}}{2}\,\alpha_1\,i\,e^{-i (\frac{k_x\,a}{2})}\,S_1} $     \\[10pt]
            \hline
          $ S_{22} =  S_{44} = \displaystyle{M\,\omega^{2} - \frac{3}{2}\,\alpha_{1}- 3\, \alpha_2+2\,\alpha_2\,C_4+\frac{\alpha_2}{2}(C_1+C_2)}$      \\[10pt]
            \hline
          $S_{23} =  S_{32}^{*} = \displaystyle{-\frac{\sqrt{3}}{2}\,\alpha_1\,i\,e^{-i (\frac{k_x\,a}{2})}\,S_1}$     \\[10pt]
            \hline
          $S_{24} = S_{42}^{*} = \displaystyle{\frac{{3}}{2}\,\alpha_1\,e^{-i (\frac{k_x\,a}{2})}\,C_3}$ \\[10pt]
            \hline
          $S_{34} = S_{43} = \displaystyle{\frac{\sqrt{3}}{2}\,\alpha_2\,(C_1-C_2)}$     \\[10pt]
          \hline
          with   $ C_1=cos (\frac{3k_xa}{2}+\frac{\sqrt{3}k_ya}{2});\ \ C_2=cos (\frac{3k_xa}{2}-\frac{\sqrt{3}k_ya}{2});$\\  $C_3=cos (\frac{\sqrt{3}k_ya}{2});\ \ C_4=cos(\sqrt{3}k_ya);\ \ S_1=sin(\frac{\sqrt{3}k_ya}{2})$ \\[10pt]
          \hline

      \end{tabular}
    \caption{Table summarising the elements of  Secular Determinant of a honeycomb lattice: (\textit{$^{*}$ represents the complex conjugate})}
    \label{table:secular det of honeycomb lattice}
  \end{table}
To validate the FFT computed results, the analytical treatment of the problem is also done by setting up the corresponding secular determinant. The equations of motion for atom $1$ as origin are derived using its neighbor details given in Table \ref{table:Reference atom and neighbors for honeycomb} and the corresponding lattice details in Table \ref{table:Hexagonal lattice details}. The differential equations of motion so formulated for atom $1$ are:\\

\begin{figure}[H]%Vibration spectrum of Hexagonal lattice
   \centering
        \includegraphics[height=5.64cm,width=8.15cm]{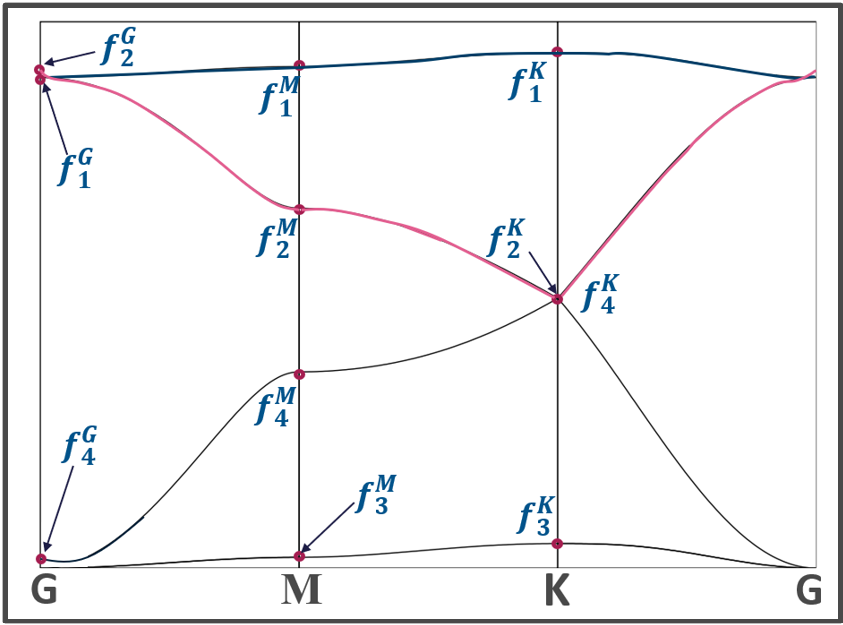}
        \caption{Phonon Spectrum ($\omega^2-k$) of honeycomb lattice in the first Brillouin zone: Temporal frequency notations are used to annotate the high symmetry points.}
        \label{fig:Vibration spectrum of Hexagon lattice}
    \end{figure}

\noindent Equation of motion in \textbf{x} direction for atom $1$:
\begin{multline}
        m \, \frac{{d^2x_{1}}}{{dt^2}} = -[\alpha_1(x_1 - x_2) - \alpha_1\left(\frac{\sqrt{3}}{4}\right)(y_4 - y_0) + \\ \alpha_1\,\left(\frac{1}{4}\right)\,(2x_1 - x_0 - x_4)  + \alpha_2\left(\frac{3}{4}\right)(4x_1 - 2x_3 - 2x_5)]
        \label{eqn:20C}
\end{multline}  
\noindent Equation of motion in \textbf{y} direction for atom $1$:
\begin{multline}
         m \, \frac{{d^2y_{1}}}{{dt^2}}= -[\alpha_1\left(\frac{3}{4}\right)(2y_1 - y_0 - y_4) - \alpha_1\left(\frac{\sqrt{3}}{4}\right)(x_4 - x_0) + \\ \alpha_2\,(2y_1 - y_3 - y_5) + \alpha_2\left(\frac{1}{4}\right)(4y_1 - 2y_3 - 2y_5)]
         \label{eqn:20D}.
\end{multline} 
\begin{figure}[H]%Brillouin zone of Hexagonal lattice
   \centering
        \includegraphics[height=5.32cm,width=6.39cm]{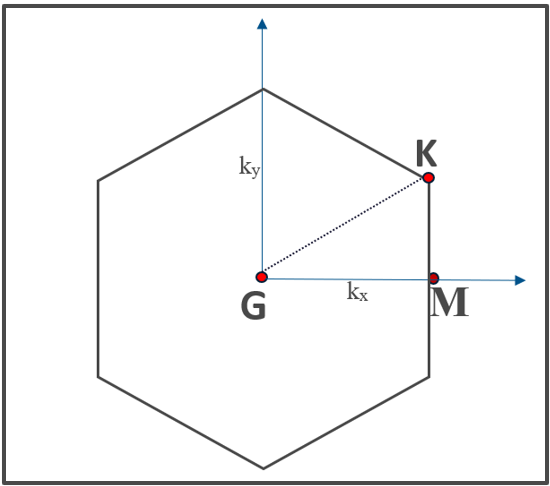}
        \caption{Brillouin zone in Hexagon reciprocal lattice with symmetry points: Annotated symmetry points are $G(0,0),\ M\left(\frac{2\pi}{3a},0
        \right),\ K\left(\frac{2\pi}{3a},\frac{2\pi}{3\sqrt{3}a}
        \right)$.}
        \label{fig:Brillouin zone of a honeycomb lattice}
    \end{figure}
The elements of the $4\times 4$ secular determinant \emph{S} obtained using Eqn.~\ref{eqn:20A} - Eqn.~\ref{eqn:20D} are summarized in Table \ref{table:secular det of honeycomb lattice}. The analytical expressions of the phonon dispersion relations obtained by solving the secular determinant at the FBZ symmetry points are given in Table~\ref{table:Phonon frequencies for honeycomb lattice}. Figure~\ref{fig:Brillouin zone of a honeycomb lattice} shows the first Brillouin zone and the location of symmetry points for a honeycomb lattice. The analytically derived phonon spectrum is depicted in Fig.~\ref{fig:Vibration spectrum of Hexagon lattice}. 
\begin{table}[h!]
    \centering
    \renewcommand{\arraystretch}{1.2} % Adjust vertical spacing
    \begin{tabular}{|>{\centering\arraybackslash}m{1.3 cm}|>{\centering\arraybackslash}m{2 cm}|>{\centering\arraybackslash}m{5.0 cm}|}
      \hline
      Phonon \linebreak Frequency & Symmetry Points  & Frequency Expression \\[12pt]
      \hline
      $f_1^{G}$ & \multirow{4}{*}{ $G\linebreak(0, 0)$}   & $\displaystyle \frac{1}{2\pi}\sqrt{\frac{3 \alpha_1}{m}}$ \\[12pt]
      $f_2^{G}$ &  & $\displaystyle \frac{1}{2\pi}\sqrt{\frac{3 \alpha_1 + 1.5 \alpha_2}{m}} $\\[  12pt]
      $f_3^{G}$ &  & $ 0 $ \\[  12pt]
      $f_4^{G}$ &  & $\displaystyle \frac{1}{2\pi}\sqrt{\frac{1.5 \alpha_2}{m}} $\\[  12pt]
      \hline
      $f_1^{M}$ & \multirow{4}{*}{$M\linebreak\left(\frac{2\,\pi}{3\,a},\ 0\right)$}    & $\displaystyle\frac{1}{2\pi}\sqrt{\frac{3 \alpha_1 + 2 \alpha_2}{m}}$ \\[  12pt]
      $f_2^{M}$ &   & $\displaystyle \frac{1}{2\pi}\sqrt{\frac{2 \alpha_1 + 6 \alpha_2}{m}}$ \\[ 12pt]
      $f_3^{M}$ &  & $\displaystyle \frac{1}{2\pi}\sqrt{\frac{2 \alpha_2}{m}}$ \\[  12pt]
      $f_4^{M}$ &  & $\displaystyle \frac{1}{2\pi}\sqrt{\frac{ \alpha_1 + 6 \alpha_2}{m}}$\\[  12pt]
      \hline
      $f_1^{K}$ & \multirow{4}{*}{$K\left( \frac{2\,\pi}{3\,a},\ \frac{2\,\pi}{3\,\sqrt{3}\,a}\right)$}   & $\displaystyle \frac{1}{2\pi}\sqrt{\frac{\frac{9}{4} \alpha_1 + 4 \alpha_2 +  \frac{1}{2}\sqrt{\frac{9}{4} \alpha_1^2 +  \alpha_2^2}}{m}}$ \\[  16pt]
      $f_2^{K}$ &  & $\displaystyle \frac{1}{2\pi}\sqrt{\frac{\frac{9}{4} \alpha_1 + 4 \alpha_2 -  \frac{1}{2}\sqrt{\frac{9}{4} \alpha_1^2 +  \alpha_2^2}}{m}} $\\[  16pt]
      $f_3^{K}$ &  & $\displaystyle\frac{1}{2\pi}\sqrt{\frac{\frac{3}{4} \alpha_1 + 4 \alpha_2 -  \frac{1}{2}\sqrt{\frac{9}{4} \alpha_1^2 +  \alpha_2^2}}{m}}$ \\[  16pt]
      $f_4^{K}$ &   & $\displaystyle \frac{1}{2\pi}\sqrt{\frac{\frac{3}{4}\alpha_1 + 4\alpha_2 +  \frac{1}{2}\sqrt{\frac{9}{4}\alpha_1^2 + \alpha_2^2}}{m}}$\\[  16pt]
      \hline
    \end{tabular}
    \caption{Table listing the analytical expressions for phonon frequencies at the high symmetry points in a monatomic honeycomb lattice in terms of nearest neighbor and next-nearest neighbor central force constants $\alpha_1$ and $\alpha_2$ respectively.}
    \label{table:Phonon frequencies for honeycomb lattice}
\end{table}

Tables~\ref{table:FFT Table for honeycomb lattice a1=4a2} and ~\ref{table:FFT Table for honeycomb lattice a1=20a2} give the mapping of analytical and computed phonon frequencies for force constant ratios $\alpha_1: \alpha_2=4:1$ and $\alpha_1: \alpha_2=20:1$, corresponding to the plots in Fig.~\ref{fig:FFT of Hexagon lattice a1=4a2} and Fig.~\ref{fig:FFT of Hexagon lattice a1=20a2} respectively. On comparing the analytical and FFT computed values of phonon frequencies, it is found that irrespective of the $\alpha_1: \alpha_2$ ratio, FFT captures the $f_1^G$ and the four \emph{K}-point frequencies in each case. The phonon branches corresponding to $f_2^K \ \&\ f_4^K$ are found to be very nearly degenerate in each case. The relative accuracy and resolution of the captured  FFT peaks, however, varies with ratio of force constants. As can be seen from the tables, the $20:1$ ratio exhibits a higher accuracy than $4:1$ ratio for all the \emph{K}-point captured peaks. The $f_1^{G'}$ and $f_1^{K'}$ peaks exhibit a better resolution for $4:1$ ratio than the $20:1$ ratio. The \emph{M}-point phonon frequencies are not captured in any of the cases except the capture of $f_4^{M'}$ peak for the $4:1$ ratio because of its coincidental close degeneracy with the $f_2^K \ \&\ f_4^K$ peaks for the given ratio of force constants. Overall, our FFT computation successfully captures the frequency peaks for each of the four normal modes of vibration expected for a monatomic honeycomb lattice. Its failure to capture the all the phonon frequencies is once again attributed to the inadequacy of our model to account for the interatomic interactions involved in the phonon modes corresponding to the missing frequencies. It is proposed that if two-body deLauney\cite{de_launey} or three-body CGW\cite{CGW} type of angular forces are used to model the nearest and next-nearest neighbors along with the central forces, one may capture the missing frequencies in the phonon spectrum of a monatomic honeycomb lattice.       
%\flushleft
 
%%%%%%%%%%%%%%%%%%%%%%%%%%%%%%%%%%%%%%%%%%%%%%%%%%%%%%%%%%%%%%%%%%%%%%%
\begin{table}[h!]
    \centering
    \renewcommand{\arraystretch}{1.2} % Adjust vertical spacing
    \begin{tabular}{|>{\centering\arraybackslash}m{2.5 cm}|>{\centering\arraybackslash}m{2.5 cm}|>{\centering\arraybackslash}m{2.5 cm}|}
      \hline
      \multicolumn{3}{|c|}{Theoretical Phonon Frequencies at the Symmetry Points} \\[7pt]
      \hline
      \large $G\,(0, 0)$ & $M\,\left(\frac{2\,\pi}{3\,a},\ 0\right)$ &$K\,\left( \frac{2\,\pi}{3\,a},\ \frac{2\,\pi}{3\,\sqrt{3}\,a}\right)$  \\[ 7pt]
       \hline
        $f_1^{G} = 12.328$ & $f_1^{M} = 13.316$ & $f_1^{K} = 14.253$ \\[5pt]
        $f_2^{G} = 13.075$ & $f_2^{M} = 13.316$ & $f_2^{K} = 11.231$ \\[5pt]
        $f_3^{G} = 0.0$ & $f_3^{M} = 5.033$ & $f_3^{K} = 7.081$ \\[5pt]
        $f_4^{G} = 4.359$ & $f_4^{M} = 11.254$ & $f_4^{K} = 11.277$ \\[5pt]
       \hline
       \multicolumn{3}{|c|}{Phonon Frequencies captured by FFT} \\[7pt]
       \hline
        $f_1^{G'} = \textcolor{red}{12.322}\linebreak (\textcolor{blue}{0.04\%})$ & $ - $ & $f_1^{K'} = \textcolor{red}{14.443}\linebreak (\textcolor{blue}{1.33\%})$ \\[5pt]
        $ - $ & $ - $ & $f_2^{K'} = \textcolor{red}{11.528}\linebreak (\textcolor{blue}{2.65\%})$ \\[5pt]
        $f_3^{G'} = \textcolor{red}{0.0}\linebreak (\textcolor{blue}{0.00\%})$ & $ - $ & $f_3^{K'} = \textcolor{red}{7.544}\linebreak (\textcolor{blue}{6.54\%})$ \\[5pt]
        $ - $ & $f_4^{M'} = \textcolor{red}{11.528}\linebreak (\textcolor{blue}{2.43\%})$ & $f_4^{K'} = \textcolor{red}{11.528}\linebreak (\textcolor{blue}{2.23\%})$ \\[5pt]
       \hline
    \end{tabular}
    \caption{Table lists the expected theoretical phonon frequency values and corresponding computational values for a honeycomb lattice with force constant ratio 4:1.}
    \label{table:FFT Table for honeycomb lattice a1=4a2}
\end{table}

\begin{table}[h!]
    \centering
    \renewcommand{\arraystretch}{1.2} % Adjust vertical spacing
    \begin{tabular}{|>{\centering\arraybackslash}m{2.5 cm}|>{\centering\arraybackslash}m{2.5 cm}|>{\centering\arraybackslash}m{2.5 cm}|}
      \hline
      \multicolumn{3}{|c|}{Theoretical Phonon Frequencies at the Symmetry Points} \\[7pt]
      \hline
      \large $G\,(0, 0)$ & $M\,\left(\frac{2\,\pi}{3\,a},\ 0\right)$ &$K\,\left( \frac{2\,\pi}{3\,a},\ \frac{2\,\pi}{3\,\sqrt{3}\,a}\right)$  \\[ 7pt]
       \hline
        $f_1^{G} = 12.328$ & $f_1^{M} = 12.531$ & $f_1^{K} = 12.733$ \\[5pt]
        $f_2^{G} = 12.482$ & $f_2^{M} = 10.794$ & $f_2^{K} = 9.279$ \\[5pt]
        $f_3^{G} = 0.0$ & $f_3^{M} = 2.250$ & $f_3^{K} = 3.179$ \\[5pt]
        $f_4^{G} = 1.949$ & $f_4^{M} =  8.115$ & $f_4^{K} = 9.281$ \\[5pt]
       \hline
       \multicolumn{3}{|c|}{Phonon Frequencies captured by FFT} \\[7pt]
       \hline
        $f_1^{G'} = \textcolor{red}{12.322}\linebreak (\textcolor{blue}{0.04\%})$ & $ - $ & $f_1^{K'} = \textcolor{red}{12.766}\linebreak (\textcolor{blue}{0.26\%})$ \\[5pt]
        $ - $ & $ - $ & $f_2^{K'} = \textcolor{red}{9.337}\linebreak (\textcolor{blue}{0.63\%})$ \\[5pt]
        $f_3^{G'} = \textcolor{red}{0.0}\linebreak (\textcolor{blue}{0.00\%})$ & $ - $ &  $f_3^{K'} = \textcolor{red}{3.368}\linebreak (\textcolor{blue}{5.94\%}) $ \\[5pt]
        $ - $ & $ - $ & $f_4^{K'} = \textcolor{red}{9.337}\linebreak (\textcolor{blue}{0.60\%})$ \\[5pt]
       \hline
    \end{tabular}
    \caption{Table lists the expected theoretical phonon frequency values and corresponding computational values for a honeycomb lattice with force constant ratio 20:1.}
    \label{table:FFT Table for honeycomb lattice a1=20a2}
\end{table}
%%%%%%%%%%%%%%%%%%%%%%%%%%%%%%%%%%%%%%%%%%%%%%%%%%%%%%%%%%%%%%%%%%%%%%%
\section{\label{sec7:}CONCLUSION\protect }
The traditional analytical method of lattice dynamical investigation assumes the existence of plane wave solutions for each atom in the unit cell. This approach relies on an implicit assumption of BvK periodic boundary conditions for constructing the secular determinant. In the pedagogical context at the undergraduate level, the method works very well for solving problems relating to 1D and 2D lattices which involve only central forces in the nearest neighbor approximation. Most of the standard solid state physics textbooks such as Omar\cite{M.Ali.Omar}, Dekker\cite{Dekker} and Kittel\cite{Kittel} often focus on analyzing simplified models for linear lattices or provide a qualitative description of phonon dynamics in simple 2D and 3D lattices assuming nearest neighbour central force inter-atomic approximations. The conventional analytical approach for an undergraduate student becomes mathematically intricate even for the relatively straightforward cases of monatomic square and simple cubic bravais lattices\cite{Ashcroft}. Our novel approach comprises of explicitly incorporating the BvK boundary conditions to condense an infinite lattice to a finite lattice, solving the equations of atomic motion in the displacement-time domain and then using the FFT technique to compute the phonon spectrum. This approach allows students to move beyond the mathematical rigor and unveil the foundational aspects of the rich physics behind periodic solids. The work serves to provide a valuable tool for visualizing lattice dynamics with the explicit implementation of PBCs, facilitating an intuitive understanding of lattice periodicity and vibrations for undergraduate students. 

The various models discussed in the present work can be easily extended to explore the lattice dynamics of other commonly encountered structures. Some exercises are suggested below for the interested reader:

\begin{enumerate}[label=\roman*]
    \item Diatomic Square lattice using central and angular forces.
    \item Hexagonal lattice with two non-equivalent lattices sites occupied by atoms of different masses.
    \item Dynamics of a simple cubic lattice, body-centred cubic lattice and face-centred cubic in the nearest neighbour central force approximation. (For reference, readers may refer to Problem 5 on Page 449-450 in Solid State Physics by Ashcroft and Mermin\cite{Ashcroft} for a face-centred cubic lattice and Pages 109-114 in the book by H.C. Gupta\cite{HC_Gupta} for body-centred cubic lattice.) The analytical expressions given in the textbooks can be used to validate and interpret the computed results of the lattices in the nearest neighbor approximation.
    \item The readers  suggested to explore the inclusion of anharmonicity in the system (for example, a cubic interatomic force as laid out in Problem 9.24 on Page 336 in the book by Harvey Gould)\cite{Gould} for the case of a 1D linear monatomic chain. The conventional linear algebra cannot be employed in cases of non-linear interacting forces. However, the approach elucidated in the paper will account for the anharmonicity in the system. 
\end{enumerate}

We firmly believe that the present work contributes to the development of a computational acumen in budding undergraduate physicists, piquing their curiosity to delve deeper into Physics through the use of numerical tools.\\

*\textit{A preliminary account of the present work was presented at the March meeting of the American Physical Society, March 20-22, 2023 [Bulletin of the American Physical Society 2023, Session TT02.00006].}

\bibliography{ref}

\begin{thebibliography}{34}%
\makeatletter
\providecommand \@ifxundefined [1]{%
 \@ifx{#1\undefined}
}%
\providecommand \@ifnum [1]{%
 \ifnum #1\expandafter \@firstoftwo
 \else \expandafter \@secondoftwo
 \fi
}%
\providecommand \@ifx [1]{%
 \ifx #1\expandafter \@firstoftwo
 \else \expandafter \@secondoftwo
 \fi
}%
\providecommand \natexlab [1]{#1}%
\providecommand \enquote  [1]{``#1''}%
\providecommand \bibnamefont  [1]{#1}%
\providecommand \bibfnamefont [1]{#1}%
\providecommand \citenamefont [1]{#1}%
\providecommand \href@noop [0]{\@secondoftwo}%
\providecommand \href [0]{\begingroup \@sanitize@url \@href}%
\providecommand \@href[1]{\@@startlink{#1}\@@href}%
\providecommand \@@href[1]{\endgroup#1\@@endlink}%
\providecommand \@sanitize@url [0]{\catcode `\\12\catcode `\$12\catcode `\&12\catcode `\#12\catcode `\^12\catcode `\_12\catcode `\%12\relax}%
\providecommand \@@startlink[1]{}%
\providecommand \@@endlink[0]{}%
\providecommand \url  [0]{\begingroup\@sanitize@url \@url }%
\providecommand \@url [1]{\endgroup\@href {#1}{\urlprefix }}%
\providecommand \urlprefix  [0]{URL }%
\providecommand \Eprint [0]{\href }%
\providecommand \doibase [0]{http://dx.doi.org/}%
\providecommand \selectlanguage [0]{\@gobble}%
\providecommand \bibinfo  [0]{\@secondoftwo}%
\providecommand \bibfield  [0]{\@secondoftwo}%
\providecommand \translation [1]{[#1]}%
\providecommand \BibitemOpen [0]{}%
\providecommand \bibitemStop [0]{}%
\providecommand \bibitemNoStop [0]{.\EOS\space}%
\providecommand \EOS [0]{\spacefactor3000\relax}%
\providecommand \BibitemShut  [1]{\csname bibitem#1\endcsname}%
\let\auto@bib@innerbib\@empty
%</preamble>
\bibitem [{\citenamefont {Yang}\ \emph {et~al.}(2019)\citenamefont {Yang}, \citenamefont {Feng}, \citenamefont {Li},\ and\ \citenamefont {Ruan}}]{Yang}%
  \BibitemOpen
  \bibfield  {author} {\bibinfo {author} {\bibfnamefont {X.}~\bibnamefont {Yang}}, \bibinfo {author} {\bibfnamefont {T.}~\bibnamefont {Feng}}, \bibinfo {author} {\bibfnamefont {J.}~\bibnamefont {Li}}, \ and\ \bibinfo {author} {\bibfnamefont {X.}~\bibnamefont {Ruan}},\ }\bibfield  {title} {\enquote {\bibinfo {title} {Stronger role of four-phonon scattering than three-phonon scattering in thermal conductivity of iii-v semiconductors at room temperature},}\ }\href {\doibase 10.1103/PhysRevB.100.245203} {\bibfield  {journal} {\bibinfo  {journal} {Phys. Rev. B}\ }\textbf {\bibinfo {volume} {100}},\ \bibinfo {pages} {245203} (\bibinfo {year} {2019})}\BibitemShut {NoStop}%
\bibitem [{\citenamefont {Omar}(1993)}]{M.Ali.Omar}%
  \BibitemOpen
  \bibfield  {author} {\bibinfo {author} {\bibfnamefont {M.}~\bibnamefont {Omar}},\ }\enquote {\bibinfo {title} {Elementary solid state physics : Principles and applications},}\ \ (\bibinfo  {publisher} {Addison-Wesley},\ \bibinfo {year} {1993})\ \bibinfo {type} {Section}~\bibinfo {chapter} {3}, pp.\ \bibinfo {pages} {68--121},\ \bibinfo {edition} {4th}\ ed.\BibitemShut {Stop}%
\bibitem [{\citenamefont {Nasrollahi}\ and\ \citenamefont {Vvedensky}(2018)}]{Nasrollahi}%
  \BibitemOpen
  \bibfield  {author} {\bibinfo {author} {\bibfnamefont {S.~H.}\ \bibnamefont {Nasrollahi}}\ and\ \bibinfo {author} {\bibfnamefont {D.~D.}\ \bibnamefont {Vvedensky}},\ }\bibfield  {title} {\enquote {\bibinfo {title} {{Local normal modes and lattice dynamics}},}\ }\href {https://doi.org/10.1063/1.5034437} {\bibfield  {journal} {\bibinfo  {journal} {Journal of Applied Physics}\ }\textbf {\bibinfo {volume} {124}} (\bibinfo {year} {2018})}\BibitemShut {NoStop}%
\bibitem [{\citenamefont {Girvin}\ and\ \citenamefont {Yang}(2019)}]{Steven}%
  \BibitemOpen
  \bibfield  {author} {\bibinfo {author} {\bibfnamefont {S.~M.}\ \bibnamefont {Girvin}}\ and\ \bibinfo {author} {\bibfnamefont {K.}~\bibnamefont {Yang}},\ }\href@noop {} {\emph {\bibinfo {title} {Modern Condensed Matter Physics}}}\ (\bibinfo  {publisher} {Cambridge University Press},\ \bibinfo {year} {2019})\ pp.\ \bibinfo {pages} {78--96}\BibitemShut {NoStop}%
\bibitem [{\citenamefont {Patterson}\ and\ \citenamefont {Bailey}(2007)}]{Patterson}%
  \BibitemOpen
  \bibfield  {author} {\bibinfo {author} {\bibfnamefont {J.}~\bibnamefont {Patterson}}\ and\ \bibinfo {author} {\bibfnamefont {B.}~\bibnamefont {Bailey}},\ }\href {\doibase 10.1007/978-3-540-34933-4} {\emph {\bibinfo {title} {Solid-State Physics: Introduction to the Theory}}},\ \bibinfo {edition} {1st}\ ed.\ (\bibinfo  {publisher} {Springer Berlin Heidelberg},\ \bibinfo {year} {2007})\ pp.\ \bibinfo {pages} {41--112}\BibitemShut {NoStop}%
\bibitem [{\citenamefont {Dekker}(1958)}]{Dekker}%
  \BibitemOpen
  \bibfield  {author} {\bibinfo {author} {\bibfnamefont {A.}~\bibnamefont {Dekker}},\ }\href {https://books.google.co.in/books?id=HpZIuAAACAAJ} {\emph {\bibinfo {title} {Solid State Physics}}},\ MacMillan student editions\ (\bibinfo  {publisher} {Macmillan},\ \bibinfo {year} {1958})\ pp.\ \bibinfo {pages} {46--51}\BibitemShut {NoStop}%
\bibitem [{\citenamefont {Ziman}(2001)}]{Ziman}%
  \BibitemOpen
  \bibfield  {author} {\bibinfo {author} {\bibfnamefont {J.}~\bibnamefont {Ziman}},\ }\href {\doibase 10.1093/acprof:oso/9780198507796.001.0001} {\emph {\bibinfo {title} {{Electrons and Phonons: The Theory of Transport Phenomena in Solids}}}}\ (\bibinfo  {publisher} {Oxford University Press},\ \bibinfo {year} {2001})\ pp.\ \bibinfo {pages} {3--49}\BibitemShut {NoStop}%
\bibitem [{\citenamefont {Gupta}(1995)}]{HC_Gupta}%
  \BibitemOpen
  \bibfield  {author} {\bibinfo {author} {\bibfnamefont {H.}~\bibnamefont {Gupta}},\ }\href@noop {} {\emph {\bibinfo {title} {Solid State Physics}}}\ (\bibinfo  {publisher} {Vikas Publishing House Pvt. Ltd},\ \bibinfo {year} {1995})\ Chap.~\bibinfo {chapter} {5}, pp.\ \bibinfo {pages} {87--118}\BibitemShut {NoStop}%
\bibitem [{\citenamefont {Kittel}(2005)}]{Kittel}%
  \BibitemOpen
  \bibfield  {author} {\bibinfo {author} {\bibfnamefont {C.}~\bibnamefont {Kittel}},\ }\href {http://www.amazon.com/Introduction-Solid-Physics-Charles-Kittel/dp/047141526X/ref=dp_ob_title_bk} {\emph {\bibinfo {title} {Introduction to Solid State Physics}}},\ \bibinfo {edition} {8th}\ ed.\ (\bibinfo  {publisher} {Wiley},\ \bibinfo {year} {2005})\ pp.\ \bibinfo {pages} {91--104}\BibitemShut {NoStop}%
\bibitem [{\citenamefont {Harrison}(1979)}]{WA.Harris}%
  \BibitemOpen
  \bibfield  {author} {\bibinfo {author} {\bibfnamefont {W.~A.}\ \bibnamefont {Harrison}},\ }\href@noop {} {\emph {\bibinfo {title} {Solid State Theory}}},\ \bibinfo {edition} {1st}\ ed.\ (\bibinfo  {publisher} {Dover Publications},\ \bibinfo {year} {1979})\ pp.\ \bibinfo {pages} {333--429}\BibitemShut {NoStop}%
\bibitem [{\citenamefont {Ashcroft}\ and\ \citenamefont {Mermin}(1976)}]{Ashcroft}%
  \BibitemOpen
  \bibfield  {author} {\bibinfo {author} {\bibfnamefont {N.~W.}\ \bibnamefont {Ashcroft}}\ and\ \bibinfo {author} {\bibfnamefont {N.~D.}\ \bibnamefont {Mermin}},\ }\href@noop {} {\emph {\bibinfo {title} {{S}olid {S}tate {P}hysics}}}\ (\bibinfo  {publisher} {Holt-Saunders},\ \bibinfo {year} {1976})\ pp.\ \bibinfo {pages} {422--450}\BibitemShut {NoStop}%
\bibitem [{\citenamefont {Kittel}(1963)}]{Kittel_Quant}%
  \BibitemOpen
  \bibfield  {author} {\bibinfo {author} {\bibfnamefont {C.}~\bibnamefont {Kittel}},\ }\href@noop {} {\emph {\bibinfo {title} {Quantum theory of solids}}}\ (\bibinfo  {publisher} {Wiley},\ \bibinfo {address} {New York},\ \bibinfo {year} {1963})\ pp.\ \bibinfo {pages} {12--48}\BibitemShut {NoStop}%
\bibitem [{\citenamefont {Levy}(1968)}]{RA_Levy}%
  \BibitemOpen
  \bibfield  {author} {\bibinfo {author} {\bibfnamefont {R.~A.}\ \bibnamefont {Levy}},\ }\href@noop {} {\emph {\bibinfo {title} {Principles of Solid State Physics}}}\ (\bibinfo  {publisher} {Academic Press New York},\ \bibinfo {year} {1968})\ pp.\ \bibinfo {pages} {74 -- 106}\BibitemShut {NoStop}%
\bibitem [{\citenamefont {{Bloch}}(1929)}]{Bloch}%
  \BibitemOpen
  \bibfield  {author} {\bibinfo {author} {\bibfnamefont {F.}~\bibnamefont {{Bloch}}},\ }\bibfield  {title} {\enquote {\bibinfo {title} {{{\"U}ber die Quantenmechanik der Elektronen in Kristallgittern}},}\ }\href {\doibase 10.1007/BF01339455} {\bibfield  {journal} {\bibinfo  {journal} {Zeitschrift fur Physik}\ }\textbf {\bibinfo {volume} {52}},\ \bibinfo {pages} {555--600} (\bibinfo {year} {1929})}\BibitemShut {NoStop}%
\bibitem [{\citenamefont {Wallis}(1964)}]{Wallis}%
  \BibitemOpen
  \bibfield  {author} {\bibinfo {author} {\bibfnamefont {R.}~\bibnamefont {Wallis}},\ }\bibfield  {title} {\enquote {\bibinfo {title} {Surface effects on lattice vibrations},}\ }\href {\doibase https://doi.org/10.1016/0039-6028(64)90053-6} {\bibfield  {journal} {\bibinfo  {journal} {Surface Science}\ }\textbf {\bibinfo {volume} {2}},\ \bibinfo {pages} {146--155} (\bibinfo {year} {1964})}\BibitemShut {NoStop}%
\bibitem [{\citenamefont {Born}\ and\ \citenamefont {Von~Karman}(1912)}]{M.Born}%
  \BibitemOpen
  \bibfield  {author} {\bibinfo {author} {\bibfnamefont {M.}~\bibnamefont {Born}}\ and\ \bibinfo {author} {\bibfnamefont {T.}~\bibnamefont {Von~Karman}},\ }\bibfield  {title} {\enquote {\bibinfo {title} {\textit{Uber Schwingungen im Raumgittern}},}\ }\href@noop {} {\bibfield  {journal} {\bibinfo  {journal} {Physikalishe Zeitschrift}\ }\textbf {\bibinfo {volume} {13}},\ \bibinfo {pages} {297--309} (\bibinfo {year} {1912})}\BibitemShut {NoStop}%
\bibitem [{\citenamefont {Kutta}(1901)}]{Kutta}%
  \BibitemOpen
  \bibfield  {author} {\bibinfo {author} {\bibfnamefont {W.}~\bibnamefont {Kutta}},\ }\bibfield  {title} {\enquote {\bibinfo {title} {Beitrag zur n\"aherungsweisen {I}ntegration totaler {D}ifferentialgleichungen},}\ }\href@noop {} {\bibfield  {journal} {\bibinfo  {journal} {Zeit. Math. Phys.}\ }\textbf {\bibinfo {volume} {46}},\ \bibinfo {pages} {435--53} (\bibinfo {year} {1901})}\BibitemShut {NoStop}%
\bibitem [{\citenamefont {Ipatova}\ \emph {et~al.}(1971)\citenamefont {Ipatova}, \citenamefont {Maradudin}, \citenamefont {Montroll},\ and\ \citenamefont {Weiss}}]{Maradudin}%
  \BibitemOpen
  \bibfield  {author} {\bibinfo {author} {\bibfnamefont {I.~P.}\ \bibnamefont {Ipatova}}, \bibinfo {author} {\bibfnamefont {A.~A.}\ \bibnamefont {Maradudin}}, \bibinfo {author} {\bibfnamefont {E.~W.}\ \bibnamefont {Montroll}}, \ and\ \bibinfo {author} {\bibfnamefont {G.~H.}\ \bibnamefont {Weiss}},\ }\href@noop {} {\emph {\bibinfo {title} {Theory of lattice dynamics in the harmonic approximation}}},\ \bibinfo {edition} {2nd}\ ed.\ (\bibinfo  {publisher} {Academic Press New York},\ \bibinfo {year} {1971})\BibitemShut {NoStop}%
\bibitem [{\citenamefont {Escande}, \citenamefont {Doveil},\ and\ \citenamefont {Elskens}(2015)}]{Escande}%
  \BibitemOpen
  \bibfield  {author} {\bibinfo {author} {\bibfnamefont {D.~F.}\ \bibnamefont {Escande}}, \bibinfo {author} {\bibfnamefont {F.}~\bibnamefont {Doveil}}, \ and\ \bibinfo {author} {\bibfnamefont {Y.}~\bibnamefont {Elskens}},\ }\bibfield  {title} {\enquote {\bibinfo {title} {N-body description of debye shielding and landau damping},}\ }\href {\doibase 10.1088/0741-3335/58/1/014040} {\bibfield  {journal} {\bibinfo  {journal} {Plasma Physics and Controlled Fusion}\ }\textbf {\bibinfo {volume} {58}},\ \bibinfo {pages} {014040} (\bibinfo {year} {2015})}\BibitemShut {NoStop}%
\bibitem [{\citenamefont {Ashdhir}\ \emph {et~al.}(2021)\citenamefont {Ashdhir}, \citenamefont {Arya}, \citenamefont {Rani},\ and\ \citenamefont {Anshika}}]{Ashdhir_etal}%
  \BibitemOpen
  \bibfield  {author} {\bibinfo {author} {\bibfnamefont {P.}~\bibnamefont {Ashdhir}}, \bibinfo {author} {\bibfnamefont {J.}~\bibnamefont {Arya}}, \bibinfo {author} {\bibfnamefont {C.~E.}\ \bibnamefont {Rani}}, \ and\ \bibinfo {author} {\bibnamefont {Anshika}},\ }\bibfield  {title} {\enquote {\bibinfo {title} {{Exploring the fundamentals of fast Fourier transform technique and its elementary applications in physics}},}\ }\href {\doibase 10.1088/1361-6404/ac20ad} {\bibfield  {journal} {\bibinfo  {journal} {European Journal of Physics}\ }\textbf {\bibinfo {volume} {42}},\ \bibinfo {eid} {065805} (\bibinfo {year} {2021})}\BibitemShut {NoStop}%
\bibitem [{\citenamefont {De~Launay}(1956)}]{de_launey}%
  \BibitemOpen
  \bibfield  {author} {\bibinfo {author} {\bibfnamefont {J.}~\bibnamefont {De~Launay}},\ }\href@noop {} {\emph {\bibinfo {title} {Solid State Physics}}},\ Vol.~\bibinfo {volume} {2}\ (\bibinfo  {publisher} {Academic Press New York},\ \bibinfo {year} {1956})\BibitemShut {NoStop}%
\bibitem [{\citenamefont {Cserti}\ and\ \citenamefont {Tichy}(2004)}]{Cserti}%
  \BibitemOpen
  \bibfield  {author} {\bibinfo {author} {\bibfnamefont {J.}~\bibnamefont {Cserti}}\ and\ \bibinfo {author} {\bibfnamefont {G.}~\bibnamefont {Tichy}},\ }\bibfield  {title} {\enquote {\bibinfo {title} {A simple model for the vibrational modes in honeycomb lattices},}\ }\href {\doibase 10.1088/0143-0807/25/6/004} {\bibfield  {journal} {\bibinfo  {journal} {European Journal of Physics}\ }\textbf {\bibinfo {volume} {25}},\ \bibinfo {pages} {723} (\bibinfo {year} {2004})}\BibitemShut {NoStop}%
\bibitem [{\citenamefont {Papoular}(2008)}]{papoular}%
  \BibitemOpen
  \bibfield  {author} {\bibinfo {author} {\bibfnamefont {D.~J.}\ \bibnamefont {Papoular}},\ }\href@noop {} {\enquote {\bibinfo {title} {Harmonic stability analysis of the 2d square and hexagonal bravais lattices for a finite--ranged repulsive pair potential. consequence for a 2d system of ultracold composite bosons},}\ } (\bibinfo {year} {2008}),\ \bibinfo {note} {{(unpublished)}},\ \Eprint {http://arxiv.org/abs/0806.4325} {arXiv:0806.4325 [cond-mat.other]} \BibitemShut {NoStop}%
\bibitem [{\citenamefont {Iachello}\ \emph {et~al.}(2015)\citenamefont {Iachello}, \citenamefont {Dietz}, \citenamefont {Miski-Oglu},\ and\ \citenamefont {Richter}}]{Iachello}%
  \BibitemOpen
  \bibfield  {author} {\bibinfo {author} {\bibfnamefont {F.}~\bibnamefont {Iachello}}, \bibinfo {author} {\bibfnamefont {B.}~\bibnamefont {Dietz}}, \bibinfo {author} {\bibfnamefont {M.}~\bibnamefont {Miski-Oglu}}, \ and\ \bibinfo {author} {\bibfnamefont {A.}~\bibnamefont {Richter}},\ }\bibfield  {title} {\enquote {\bibinfo {title} {Algebraic theory of crystal vibrations: Singularities and zeros in vibrations of one- and two-dimensional lattices},}\ }\href {\doibase 10.1103/PhysRevB.91.214307} {\bibfield  {journal} {\bibinfo  {journal} {Phys. Rev. B}\ }\textbf {\bibinfo {volume} {91}},\ \bibinfo {pages} {214307} (\bibinfo {year} {2015})}\BibitemShut {NoStop}%
\bibitem [{\citenamefont {Lang}\ \emph {et~al.}(1994)\citenamefont {Lang}, \citenamefont {Doyen-Lang}, \citenamefont {Charlier},\ and\ \citenamefont {Charlier}}]{lang}%
  \BibitemOpen
  \bibfield  {author} {\bibinfo {author} {\bibfnamefont {L.}~\bibnamefont {Lang}}, \bibinfo {author} {\bibfnamefont {S.}~\bibnamefont {Doyen-Lang}}, \bibinfo {author} {\bibfnamefont {A.}~\bibnamefont {Charlier}}, \ and\ \bibinfo {author} {\bibfnamefont {M.~F.}\ \bibnamefont {Charlier}},\ }\bibfield  {title} {\enquote {\bibinfo {title} {Dynamical study of graphite and graphite intercalation compounds},}\ }\href {\doibase 10.1103/PhysRevB.49.5672} {\bibfield  {journal} {\bibinfo  {journal} {Phys. Rev. B}\ }\textbf {\bibinfo {volume} {49}},\ \bibinfo {pages} {5672--5681} (\bibinfo {year} {1994})}\BibitemShut {NoStop}%
\bibitem [{\citenamefont {Maeda}, \citenamefont {Kuramoto},\ and\ \citenamefont {Horie}(1979)}]{Maeda}%
  \BibitemOpen
  \bibfield  {author} {\bibinfo {author} {\bibfnamefont {M.}~\bibnamefont {Maeda}}, \bibinfo {author} {\bibfnamefont {Y.}~\bibnamefont {Kuramoto}}, \ and\ \bibinfo {author} {\bibfnamefont {C.}~\bibnamefont {Horie}},\ }\bibfield  {title} {\enquote {\bibinfo {title} {Phonon dispersion relations of graphite},}\ }\href {\doibase 10.1143/JPSJ.47.337} {\bibfield  {journal} {\bibinfo  {journal} {Journal of the Physical Society of Japan}\ }\textbf {\bibinfo {volume} {47}},\ \bibinfo {pages} {337--338} (\bibinfo {year} {1979})}\BibitemShut {NoStop}%
\bibitem [{\citenamefont {Gupta}\ \emph {et~al.}(1986)\citenamefont {Gupta}, \citenamefont {Malhotra}, \citenamefont {Rani},\ and\ \citenamefont {Tripathi}}]{Gupta_malhotra}%
  \BibitemOpen
  \bibfield  {author} {\bibinfo {author} {\bibfnamefont {H.~C.}\ \bibnamefont {Gupta}}, \bibinfo {author} {\bibfnamefont {J.}~\bibnamefont {Malhotra}}, \bibinfo {author} {\bibfnamefont {N.}~\bibnamefont {Rani}}, \ and\ \bibinfo {author} {\bibfnamefont {B.~B.}\ \bibnamefont {Tripathi}},\ }\bibfield  {title} {\enquote {\bibinfo {title} {Lattice-dynamical model for graphite and its alkali-metal intercalation compounds},}\ }\href {\doibase 10.1103/PhysRevB.33.7285} {\bibfield  {journal} {\bibinfo  {journal} {Phys. Rev. B}\ }\textbf {\bibinfo {volume} {33}},\ \bibinfo {pages} {7285--7287} (\bibinfo {year} {1986})}\BibitemShut {NoStop}%
\bibitem [{\citenamefont {Woods}\ and\ \citenamefont {Mahan}(2000)}]{Woods}%
  \BibitemOpen
  \bibfield  {author} {\bibinfo {author} {\bibfnamefont {L.~M.}\ \bibnamefont {Woods}}\ and\ \bibinfo {author} {\bibfnamefont {G.~D.}\ \bibnamefont {Mahan}},\ }\bibfield  {title} {\enquote {\bibinfo {title} {Electron-phonon effects in graphene and armchair (10,10) single-wall carbon nanotubes},}\ }\href {\doibase 10.1103/PhysRevB.61.10651} {\bibfield  {journal} {\bibinfo  {journal} {Phys. Rev. B}\ }\textbf {\bibinfo {volume} {61}},\ \bibinfo {pages} {10651--10663} (\bibinfo {year} {2000})}\BibitemShut {NoStop}%
\bibitem [{\citenamefont {Damnjanović}\ \emph {et~al.}(2003)\citenamefont {Damnjanović}, \citenamefont {Dobardžić}, \citenamefont {Milošević}, \citenamefont {Vuković},\ and\ \citenamefont {Nikolić}}]{Damnjanovic}%
  \BibitemOpen
  \bibfield  {author} {\bibinfo {author} {\bibfnamefont {M.}~\bibnamefont {Damnjanović}}, \bibinfo {author} {\bibfnamefont {E.}~\bibnamefont {Dobardžić}}, \bibinfo {author} {\bibfnamefont {I.}~\bibnamefont {Milošević}}, \bibinfo {author} {\bibfnamefont {T.}~\bibnamefont {Vuković}}, \ and\ \bibinfo {author} {\bibfnamefont {B.}~\bibnamefont {Nikolić}},\ }\bibfield  {title} {\enquote {\bibinfo {title} {Lattice dynamics and symmetry of double wall carbon nanotubes},}\ }\href {\doibase 10.1088/1367-2630/5/1/148} {\bibfield  {journal} {\bibinfo  {journal} {New Journal of Physics}\ }\textbf {\bibinfo {volume} {5}},\ \bibinfo {pages} {148} (\bibinfo {year} {2003})}\BibitemShut {NoStop}%
\bibitem [{\citenamefont {Jishi}\ \emph {et~al.}(1993)\citenamefont {Jishi}, \citenamefont {Venkataraman}, \citenamefont {Dresselhaus},\ and\ \citenamefont {Dresselhaus}}]{Jishi}%
  \BibitemOpen
  \bibfield  {author} {\bibinfo {author} {\bibfnamefont {R.}~\bibnamefont {Jishi}}, \bibinfo {author} {\bibfnamefont {L.}~\bibnamefont {Venkataraman}}, \bibinfo {author} {\bibfnamefont {M.}~\bibnamefont {Dresselhaus}}, \ and\ \bibinfo {author} {\bibfnamefont {G.}~\bibnamefont {Dresselhaus}},\ }\bibfield  {title} {\enquote {\bibinfo {title} {Phonon modes in carbon nanotubules},}\ }\href {\doibase https://doi.org/10.1016/0009-2614(93)87205-H} {\bibfield  {journal} {\bibinfo  {journal} {Chemical Physics Letters}\ }\textbf {\bibinfo {volume} {209}},\ \bibinfo {pages} {77--82} (\bibinfo {year} {1993})}\BibitemShut {NoStop}%
\bibitem [{\citenamefont {Mousavi}, \citenamefont {Nafisi},\ and\ \citenamefont {Maibach}(2017)}]{Mousavi}%
  \BibitemOpen
  \bibfield  {author} {\bibinfo {author} {\bibfnamefont {S.~Z.}\ \bibnamefont {Mousavi}}, \bibinfo {author} {\bibfnamefont {S.}~\bibnamefont {Nafisi}}, \ and\ \bibinfo {author} {\bibfnamefont {H.~I.}\ \bibnamefont {Maibach}},\ }\bibfield  {title} {\enquote {\bibinfo {title} {Fullerene nanoparticle in dermatological and cosmetic applications},}\ }\href {\doibase https://doi.org/10.1016/j.nano.2016.10.002} {\bibfield  {journal} {\bibinfo  {journal} {Nanomedicine: Nanotechnology, Biology and Medicine}\ }\textbf {\bibinfo {volume} {13}},\ \bibinfo {pages} {1071--1087} (\bibinfo {year} {2017})}\BibitemShut {NoStop}%
\bibitem [{\citenamefont {Matija}\ \emph {et~al.}(2013)\citenamefont {Matija}, \citenamefont {Tsenkova}, \citenamefont {Mun{\'{c}}an}, \citenamefont {Miyazaki}, \citenamefont {Banba}, \citenamefont {Tomi{\'{c}}},\ and\ \citenamefont {Jefti{\'{c}}}}]{Matija}%
  \BibitemOpen
  \bibfield  {author} {\bibinfo {author} {\bibfnamefont {L.}~\bibnamefont {Matija}}, \bibinfo {author} {\bibfnamefont {R.}~\bibnamefont {Tsenkova}}, \bibinfo {author} {\bibfnamefont {J.}~\bibnamefont {Mun{\'{c}}an}}, \bibinfo {author} {\bibfnamefont {M.}~\bibnamefont {Miyazaki}}, \bibinfo {author} {\bibfnamefont {K.}~\bibnamefont {Banba}}, \bibinfo {author} {\bibfnamefont {M.}~\bibnamefont {Tomi{\'{c}}}}, \ and\ \bibinfo {author} {\bibfnamefont {B.}~\bibnamefont {Jefti{\'{c}}}},\ }\bibfield  {title} {\enquote {\bibinfo {title} {Fullerene based nanomaterials for biomedical applications: Engineering, functionalization and characterization},}\ }\href {\doibase 10.4028/www.scientific.net/AMR.633.224} {\bibfield  {journal} {\bibinfo  {journal} {Advanced Materials Research}\ }\textbf {\bibinfo {volume} {633}},\ \bibinfo {pages} {224--238} (\bibinfo {year} {2013})}\BibitemShut {NoStop}%
\bibitem [{\citenamefont {Clark}, \citenamefont {Gazis},\ and\ \citenamefont {Wallis}(1964)}]{CGW}%
  \BibitemOpen
  \bibfield  {author} {\bibinfo {author} {\bibfnamefont {B.~C.}\ \bibnamefont {Clark}}, \bibinfo {author} {\bibfnamefont {D.~C.}\ \bibnamefont {Gazis}}, \ and\ \bibinfo {author} {\bibfnamefont {R.~F.}\ \bibnamefont {Wallis}},\ }\bibfield  {title} {\enquote {\bibinfo {title} {Frequency spectra of body-centered cubic lattices},}\ }\href {\doibase 10.1103/PhysRev.134.A1486} {\bibfield  {journal} {\bibinfo  {journal} {Phys. Rev.}\ }\textbf {\bibinfo {volume} {134}},\ \bibinfo {pages} {A1486--A1491} (\bibinfo {year} {1964})}\BibitemShut {NoStop}%
\bibitem [{\citenamefont {Gould}, \citenamefont {Tobochnik},\ and\ \citenamefont {Christian}(2007)}]{Gould}%
  \BibitemOpen
  \bibfield  {author} {\bibinfo {author} {\bibfnamefont {H.}~\bibnamefont {Gould}}, \bibinfo {author} {\bibfnamefont {J.}~\bibnamefont {Tobochnik}}, \ and\ \bibinfo {author} {\bibfnamefont {W.}~\bibnamefont {Christian}},\ }\href@noop {} {\emph {\bibinfo {title} {An Introduction to Computer Simulation Methods Third Edition (revised)}}},\ \bibinfo {edition} {3rd}\ ed.\ (\bibinfo {year} {2007})\ p.\ \bibinfo {pages} {336}\BibitemShut {NoStop}%
\end{thebibliography}%

\end{document}